\documentclass[amsmath,amssymb,twocolumn,superscriptaddress]{revtex4}

\usepackage{graphicx}
\usepackage{dcolumn}

\def\be{\begin{eqnarray}}
\def\ee{\end{eqnarray}}
\def\beq{\begin{eqnarray}}
\def\eeq{\end{eqnarray}}

\usepackage{bm}
\usepackage{graphicx}
\usepackage{dcolumn}
\usepackage{amsmath}
\usepackage[latin1]{inputenc}
\usepackage{graphicx, psfrag}
\usepackage{amssymb}
\usepackage[colorlinks=true, citecolor=blue, urlcolor = blue, linkcolor= red, bookmarks=true]{hyperref}
\usepackage{float}
\usepackage{amsmath}
\usepackage{amsfonts}
\usepackage{dcolumn}
\usepackage{hyperref}
\usepackage{subfigure}
\usepackage{pgfplots}
\usepackage{epstopdf}
\usepackage{booktabs}
\usepackage{multirow}

\begin{document}


\title{Impact of energy-momentum conservation violation on the configuration of compact stars  and their GW echoes}  


\author{Jyatsnasree Bora}
\email[Email:]{ jyatnasree.borah@gmail.com}
\affiliation{Department of Mathematical and Physical Sciences, 
College of Arts and Sciences, University of Nizwa, Nizwa, Sultanate of Oman}

\author{Dhruba Jyoti Gogoi}
\email[Email:]{moloydhruba@yahoo.in}
\affiliation{Department of Physics, Moran College, Moranhat, Charaideo 785670, Assam, India}

\author{S. K. Maurya}
\email[Email:]{sunil@unizwa.edu.om}
\affiliation{Department of Mathematical and Physical Sciences,
College of Arts and Sciences, University of Nizwa 616, Nizwa, Sultanate of Oman}

\author{Ghulam Mustafa}
\email[Email:]{gmustafa3828@zjnu.edu.cn}
\affiliation{Department of Physics Zhejiang Normal University Jinhua 321004, People's Republic of China.}

\date{\today}

\begin{abstract}
This work investigates the impacts of energy-momentum conservation violation on the configuration of strange stars constraint with gravitational wave (GW) event  GW190814 as well as eight recent observations of compact objects. The GW echoes from these interesting classes of compact objects are also calculated. To describe the matter of strange stars, we have used two different equations of state (EoSs): first an ad-hoc exotic EoS, the stiffer MIT Bag model and next realistic CFL phase of quark matter EoS. We choose Rastall gravity as a simple model with energy-momentum conservation violation with a set of model parameter values.
Our results show that this gravity theory permits stable solutions of strange stars and the resulting structures can foster GW echoes. We illustrate the implication of the gravity theory and found that the negative values of the Rastall parameter result in more compact stellar configurations and lower GW echo frequency. With an increase in the Rastall parameter, both the compactness of the stellar configurations and echo time decrease. It is worth mentioning here that with the chosen set of some probable strange star candidates from observational data and also in light of GW 190814, we have evaluated the radii of stellar models. Also, the GW echo frequencies associated with strange stars are found to be in the range of {$\approx 9-27$ kHz} for both cases. {From this work, it is also inferred that the assumption regarding the equivalence of Rastall's theory to Einstein's theory is refuted as we have noticed many deviations in the physical properties of the considered compact stars.}\\
\textbf{Keywords}: modified gravity; neutron stars; astrophysical compact stars; GW echoes
\end{abstract}

\maketitle

\section{Introduction}\label{sec01}
The current understanding of the physics of compact objects and black holes have reached a definite destination after the observations of gravitational wave (GW) from compact binary mergers by LIGO/Virgo collaborations  \cite{prl116,prl118,prl119,prx9,gw190814}. These compact binary mergers have inspired physicists to think about other exotic compact objects. The exotic compact objects maintain their identity like a black hole with high compactness and without the presence of an event horizon \cite{Zhang,Zhang2,jb4}. Among the various distinctive and interesting features of these exotic compact objects, one feature that has gained significant importance presently is the reflection of GWs from their stellar surface or GW echoes \cite{Zhang,Zhang2,jb4,2018_Pani,2018_manarelli,2019_Urbano,jb1,jb2,jb3}. Such reflection happens when a wave falls inside the gravitational potential barrier of the object, they travel to a reflecting boundary before returning to the barrier at the photon sphere after some time delay \cite{Zhang,Zhang2}. Thus it is interesting to search for such features of these compact objects. 
Furthermore, through GW and electromagnetic spectrum measurements, we have recently learned a lot about compact stellar structures. These observations' analysis imposed restrictions on macroscopic characteristics including mass, radius, and tidal deformability. We were able to constrain the composition of this matter thanks to NICER's (Neutron star Interior Composition ExploreR) observations of two compact objects (PSR J0030+0451 and PSR J0740+6620) \cite{0451,0451_2,6620_1,6620_2}. The mass-radius constraints obtained for these candidates from NICER aid in our understanding of the characteristics of compact stars \cite{0451,0451_2,6620_1,6620_2}. Furthermore, recent comments from the observation of PSR J0952-0607 reveal the lower limit of maximum attainable mass by compact stars is $M=2.35\pm0.17M_{\odot}$ \cite{romani}. It is the heaviest pulsar observed to date \cite{romani}. The GW event GW190814 is another important event to study compact objects. According to this signal, a black hole and a compact star with masses between 2.5 and 2.67$M_\odot$ coalesce in binary form. This secondary component mass range is located in the mass gap region (where it may be a compact star or black hole) \cite{gw190814}. The observations of pulsars like PSR J1614-2230 and PSR J0348+0432 have set an upper bound on the mass of $M=2M_\odot$ \cite{Tangphati1}.

The explanations of gravitational phenomena of the present universe are based on Einstein's revolutionary theory of gravitation: general relativity (GR). This theory is found to be adequate in both the short and long ranges. However, from the present perspective of the accelerated expansion of the universe, it requires the presence of some exotic dark staffs known as dark matter (DM) and Dark energy (DE). With the introduction of these dark staffs, GR can pass the test of higher energy scales. However, being a classical theory it could yield some singularity at high energy scales. By including the quantum corrections such singularities can be resolved. Again, supporting experimental results for the DE and DM are still lacking. So, alternatives to GR are a fruitful way to go beyond these issues \cite{gogoi1,gogoi_cosmo,gogoi4, Koussour23A, Koussour23B}. Recently, several studies related to compact objects and exotic objects, including wormholes have been done in different modified theories \cite{jb2,sk19,gogoi5,23_sk,Lambiase23,GogoifQ, Sekhmani23,gogoi3,gogoi2,2021_Silva,2021_Salako,2021_Ghosh,2019_Maulana,Astash01,Astash1}. Among the various proposed modifications of Einstein's GR, Rastall's theory of gravity is one that describes the evolution of astrophysical models which are non-conservative in nature \cite{rastall}. The essence of Rastall gravity is associated with strong gravity regimes \cite{23_Gogoi2, gogoi3}. Thus it is interesting to find its consequences in large curvature environments like black holes, neutron stars, strange stars and other possible exotic compact objects. 

In this work, we want to explore the less visited compact object, called the strange stars. The concept of strange quark matter or strange matter was theoretically postulated around half a century earlier by Bodmer and Witten independently \cite{1971_Bodmer,1984_Witten}. It has been suggested that such forms of matter can act as the strong interaction's ground state and are considered as a fluid with equal numbers of deconfined u, d and s quarks \cite{1986_Alcock}. To provide electrical charge neutrality smaller numbers of electrons are also present in this matter. Compact objects formed of such quark matter are known as strange quark stars or strange stars. These are hypothetical types of compact stars that are believed to be composed of strange matter. The concept of strange stars was first proposed in 1986 by Alcock, Farhi, and Olinto \cite{1986_Alcock} as a possible explanation for the high-mass compact objects observed in X-ray binaries, and has since been the subject of extensive theoretical and observational studies. Strange stars are believed to be much denser than neutron stars, with typical densities on the order of several times that of atomic nuclei \cite{Weber1999}. One of the key features of strange stars is their unique EoS, which describes the relationship between pressure, density, and temperature. {Unlike neutron stars, which are thought to be composed of dense, cold nucleons, strange stars are believed to have higher compactness, with higher central densities and maximum masses} \cite{1986_Alcock}. This has important implications for our understanding of the properties and evolution of compact stars, as well as for the study of GWs and their sources \cite{Weber1999}. The structural behaviour and properties of these compact objects is a burgeoning field of exploration. The studies on the echoes of GWs are also a topic of great interest in the current scenario \cite{2018_Pani,2018_manarelli,2019_Urbano,Zhang,jb1,jb2,jb3}. Such echoes from strange stars have been proposed as a potential signature of strange matter \cite{2018_manarelli}. According to this scenario, the surface of a strange star would behave as a reflective membrane, reflecting the incoming GW
back into the interior of the star. This would result in the generation of a series of secondary echoes, which would contain unique information about the properties of the strange matter and the EoS of the star. The detection of such echoes would provide strong evidence for the existence of strange stars and would have important implications for our understanding of the nature of matter at extremely high densities \cite{2018_manarelli}. From the observational point, though several compact objects are conjectured to be strange quark star, yet questions on their existence remains open.
One may note that the detectability of GW echoes from strange stars is still a topic of active research and debate, with some studies suggesting that the echoes may be too weak to be observable with current or near-future GW detectors \cite{2018_Pani,2018_manarelli,jb1,jb2,jb4}. Nevertheless, the possibility of detecting GW echoes from strange stars has motivated ongoing efforts to improve the sensitivity of GW detectors and to develop new data analysis techniques for the extraction of echoes from noisy data \cite{2018_manarelli,jb1,jb2,jb4}.

The literature survey confirms that the possibilities of echoes of GWs are presented mainly in the general relativistic regimes \cite{2018_Pani,2018_manarelli,Zhang,jb1,jb2}. At the same time, such possibilities in the regime of alternative gravity theory are another promising area that needs more attention. Such studies on echoes from ultracompact objects in $f(R)$ and $f(R,T)$ gravity are reported in Ref. \cite{jb3,jb4}. An important and exciting alternative of GR which has gained significant importance in recent days' research is the so-called Rastall theory of gravity \cite{rastall}. Rastall's proposal addresses the conservation law of the energy-momentum tensor, i.e., $(\nabla_{\upsilon}T^{\epsilon\upsilon}=0)$ in a curved space-time, which is a key factor in his modification of GR. Rastall's proposal suggests that the covariant divergence of the energy-momentum tensor is directly proportional to the derivative of the scalar curvature, i.e., $(\nabla_{\upsilon}T^{\epsilon\upsilon} \propto R^{;\upsilon})$. A coupling parameter was introduced for this modification, whose particular limiting value (i.e., zero coupling) lapses back this modified form to the GR. Additionally, Rastall's theory has simpler field equations compared to other modified theories, making it easier to investigate. Thus the essence of Rastall gravity due to the coupling parameter is associated with a strong gravity regime. Thus it is interesting to find its consequences in such large curvature environments. 
{Rastall theory of gravity couples the geometry to the matter fields in a non-minimal way \cite{Moradpour17,morad}, however, it has simpler field equations than those of the curvature-matter theory \cite{Koivisto,Bertolami,Harko}. It also shows good agreement with some observational data on the Hubble parameter and the age of the universe \cite{rawaf1}. Furthermore, it has better consistency with the observational data of the matter-dominated era against the Einstein field equations \cite{rawaf2,Li}. There are a plethora of articles which demand its consistency with various cosmic eras \cite{rawaf3,Majernik,Arbab}. Moreover, it is worth mentioning that Rastall's theory works as a great platform to study the gravitational lensing \cite{Rahman1,Rahman2,ziaie}. One may note that the Lagrangian of the curvature-matter theory of gravity differs from that of the Rastall theory \cite{Moradpour17}. So it is useful to investigate the structure of compact stars and their properties in the framework of Rastall gravity in order to get new aspects of these compact objects. Unlike the case of Einstein GTR, Rastall gravity considers a non-divergence-free energy-momentum. 
In 2018, M. Visser claimed on the equivalence of the both theories \cite{visser}. After that Darabi {\it et al.} compared both theories and argued that the Rastall theory is more general than the GR \cite{Darabi}. It is reported therein that the Rastall theory is an "open" theory when contrasted with the GR. A detailed discussion on the non-equivalence of GTR with Rastall theory can be found in the article \cite{Darabi}. Regarding the viability of Rastall theory and its generalization one may refer to the article \cite{2015_Oliveira}.}
In the context of neutron stars, Rastall's theory of gravity is considered in the article \cite{2015_Oliveira}. For some realistic EoSs to describe neutron star's interior, the authors have presented the constraints on the deviations of GR caused by Rastall gravity. The radial pulsations of such hypothetical stars in Rastall gravity are earlier reported in Ref. \cite{2019_Maulana}. Studies on strange star models from Rastall gravity are also reported recently in Ref. \cite{2023_Banerjee}. Considering the MIT Bag model EoS, the authors have investigated the impact of the Rastall parameter on the different physical properties of stars. Charged strange stars in this theory are investigated earlier in article \cite{2022_mustafa}. In Rastall gravity via embedding approach, compact stellar structures using a linear form of EoS are reported to investigate in Ref. \cite{2021_mustafa}. In this theoretical framework, other important studies on compact stars can be found in the Refs. \cite{zubair,mustafa,abbas,abbas2}. Besides strange stars, other highly dense objects are also recently explored in this gravity theory \cite{20_sk,21_gogoi,23_gogoi,23_Gogoi2}. Charged anisotropic strange stars using MIT Bag model in Rastall gravity are investigated in the Ref. \cite{2019_Shahzad}. Furthermore, recent works on such hypothetical objects have been able to draw significant attention in the last few years \cite{22_sk}. For compact stars using different recent astrophysical observations are reported recently in Ref. \cite{2023_kumar}.

{Motivated from the above-discussed works,} in this current work{, we want to study the} emittance of echoes of GWs in Rastall's modification to gravity. In this fascinating theory of gravity, such types of possibilities from ultracompact objects like strange stars are not discussed earlier. Besides this exciting property, the current work addresses a detailed analysis of compact stellar structures and their stabilities. {Moreover in this work we are also motivated to this study is to examine the claim of M. Visser regarding the equivalence of Rastall theory of gravity and the Einstein equations \cite{visser}. In this regard, we are going to compare the results in Rastall theory and general relativity.}
The present paper is organized after this introduction as: in the next section (Section \ref{sec02}), we have discussed Rastall gravity in brief. In Section \ref{sec03}, we have discussed the equations of the state of compact stars. The hydrostatic equilibrium equations are also discussed in the context of Rastall gravity. In Section \ref{sec03}, the physical properties of stellar models are added. For this physical analysis and stability considerations, we have discussed the mass-radius profiles, surface-redshift, and relativistic adiabatic index. In Section \ref{sec05}, we have studied echoes of GWs associated with the star models. Finally, in Section \ref{sec06}, we have summarised our results with a brief conclusion. {Throughout this work we have considered $G=c=1$.}

\section{The Rastall's approach of gravity}\label{sec02}
To describe stellar structures with spherically symmetric matter distribution in a static geometry, we consider spacetime as 
\begin{eqnarray}\label{eq1}
ds^2 = B(r)dt^2-A(r)dr^2-r^2(d\theta^2+\sin^2\theta d\phi^2).
\end{eqnarray}
For the sake of simplicity, the effect of the magnetic field and rotational frequency of the matter content of an ideal fluid are neglected here. In such a scenario, the  energy-momentum tensor can be given as
\begin{eqnarray}\label{eq2}
T_{\epsilon \upsilon} &=& (\rho + p) u_\epsilon u_\upsilon - p g_{\epsilon\upsilon}.
\end{eqnarray}
In this expression $u_\alpha$ is the four velocity and $u_\epsilon u^{\epsilon} = 1$. $g_{\epsilon\upsilon}$ represents the component from metric tensor. 
With this energy-momentum tensor \ref{eq2} and the metric \ref{eq1}, the Einstein field equations can be used to know the hydrostatic equilibrium equations of a star. Such hydrostatic equations are popularly known as the Tolman-Oppenheimer-Volkoff (TOV) equations. While deriving these equations, an important point is the conservation law of the form
\begin{eqnarray}\label{eq3}
T^{\epsilon\upsilon}_{;\epsilon}=0,
\end{eqnarray}
this equation is concurrent with the use of Einstein field equations,
\begin{eqnarray}\label{eq4}
R_{\epsilon\upsilon} - \frac{1}{2}g_{\epsilon\upsilon} R = 8\pi T_{\epsilon\upsilon},
\end{eqnarray}
where $R_{\epsilon\upsilon}$ is the Ricci tensor and $R$ is the Ricci scalar. Using these equations one can obtain the TOV equations in GR as \cite{tolman,oppenheimer}
\begin{eqnarray}\label{eq5}
\frac{dp}{dr} &=& -\frac{m \rho}{r^2}\frac{\left(1+\frac{p}{\rho}\right)\left(1+\frac{4\pi r^3 p}{m}\right)}{1 - \frac{2m}{r}}\quad,
\end{eqnarray}
and
\begin{eqnarray}\label{eq6}
\frac{dm}{dr} &=& 4\pi \rho r^2 dr.
\end{eqnarray}   
For a detailed revision of this equation, one can follow standard text like \cite{weinberg1972gravitation, glendenning2000compact}.

{For the case of Rastall gravity, the conservation law becomes \cite{rastall,Moradpour16,Moradpour17},}
\begin{eqnarray}\label{eq7}
T^{\epsilon\upsilon}_{;\epsilon} = \alpha R^{,\upsilon}. 
\end{eqnarray}
{Here, $\alpha$ denotes the Rastall constant parameter and we remark that when $\alpha \rightarrow 0$, one regains the standard GR theory. The field equation according to Rastall's proposal \cite{rastall,Moradpour16,Moradpour17} can be given as}
\begin{eqnarray}\label{eq8}
R_{\epsilon\upsilon} - \frac{1}{2}g_{\epsilon\upsilon} R +k\alpha g_{\epsilon\upsilon} R&=& k T_{\epsilon\upsilon}\nonumber\\ or, G_{\epsilon\upsilon} +k\alpha g_{\epsilon\upsilon} R &=& k T_{\epsilon\upsilon},
\end{eqnarray}
{here $k$ and $\alpha$ represent Rastall's parameter and Rastall's gravitational coupling constant respectively. Also, $k=\delta k_E$ \cite{2021_Salako,hanafy,hanafy2}, here the constant $\delta$ is an arbitrary parameter defined as $\delta=\dfrac{4\gamma-1}{6\gamma-1}$ and standard GR has its value 1. $k_E=8\pi$ is Einstein gravitational coupling constant. 
The Newtonian limit of the Rastall theory implies that}
\begin{eqnarray}\label{eq8_2}
k=\dfrac{\gamma}{\alpha}=\dfrac{4\gamma-1}{6\gamma-1}8\pi 
\end{eqnarray}
{The trace of the above equation \eqref{eq8} provides,}
\begin{eqnarray}\nonumber
R(4k\alpha-1)=T.
\end{eqnarray}
{Now, as $T$ is not always zero, thus $k\alpha=1/4$ is not
an allowed value in this theory \cite{rastall}. Equation \eqref{eq8}, for $\alpha=0$ will corresponds to the GR equation \eqref{eq4}. These field equations \eqref{eq8} may be rewritten in the form by considering $\alpha=\dfrac{1}{2k}(1-\lambda)$ \cite{rastall,2015_Oliveira,2019_Maulana,Tangphati,Banerjee23} as,}
\begin{eqnarray}\label{eq8_1}
R_{\epsilon\upsilon} - \frac{\lambda}{2}g_{\epsilon\upsilon} R = k T_{\epsilon\upsilon}. 
\end{eqnarray}
{From relation \eqref{eq8_2}, it is obvious that only for $\lambda=\alpha=0$, we have $k=8\pi$. In 2015, Oliveira et al. first obtained the modified hydrostatic equilibrium equation for compact stars in Rastall gravity \cite{2015_Oliveira}. The authors have studied neutron stars in this theory considering $k=8\pi$ \cite{2015_Oliveira}. The authors have reported that the value of $\alpha$ is close to zero. It is also conferred that consideration of $k=8\pi$ has not rejected the Rastall hypothesis \cite{2015_Oliveira, 2017_Moradpour}. With this consideration, several studies are going on to study compact objects in the realm of Rastall gravity \cite{2019_Maulana,Tangphati,Banerjee23}.}

{With this consideration, one can write the field equation for Rastall gravity \eqref{eq8_1} as \cite{2015_Oliveira},}
\begin{eqnarray}\label{eq8_3}
R_{\epsilon\upsilon} - \frac{\lambda}{2}g_{\epsilon\upsilon} R = \kappa T_{\epsilon\upsilon}. 
\end{eqnarray}

{ The components are:
\begin{eqnarray}
     \label{com00}
&&\hspace{-0.6cm} \kappa \rho(r)= \frac{1}{{4 r^2 A(r)^2 B(r)^2}}\Big[r B(r) A'(r) \left((\lambda -1) r B'(r)+4 \lambda  B(r)\right)\nonumber\\&&\hspace{0.5cm}+A(r) \Big((\lambda -1) r^2 B'(r)^2-2 (\lambda -1) r B(r) \big(r B''(r)\nonumber\\&&\hspace{0.5cm}+2 B'(r)\big)-4 \lambda  B(r)^2\Big)+4 \lambda  A(r)^2 B(r)^2\Big],\\\label{com11}
&&\hspace{-0.6cm} \kappa p(r) =\frac{-1}{{4 r^2 A(r)^2 B(r)^2}}  \Big[(\lambda -1) r B(r) A'(r) \left(r B'(r)+4 B(r)\right)\nonumber\\&&\hspace{0.5cm}+A(r) \Big((\lambda -1) r^2 B'(r)^2+2 r B(r) \big(-(\lambda -1) r B''(r)\nonumber\\&&\hspace{0.5cm}-2 \lambda  B'(r)\big)-4 \lambda  B(r)^2\Big)+4 \lambda  A(r)^2 B(r)^2\Big],\\\label{com22}
&&\hspace{-0.6cm} \kappa p(r) = \frac{-1}{{4 r^2 A(r)^2 B(r)^2}}\Big[r B(r) A'(r) \left(\lambda  r B'(r)+(4 \lambda -2) B(r)\right)\nonumber\\&&\hspace{0.5cm}+A(r) \Big(\lambda  r^2 B'(r)^2-2 r B(r) \big(\lambda  r B''(r)+(2 \lambda -1) B'(r)\big)\nonumber\\&&\hspace{0.5cm}-4 (\lambda -1) B(r)^2\Big)+4 (\lambda -1) A(r)^2 B(r)^2\Big]. 
\end{eqnarray}}

{
The above components of the field equations or TOV equations can be expressed in a GR like form by using a parametric redefinition:

\begin{eqnarray}\label{eq11}
\tilde{x}&=& \alpha_1 \rho + 3\alpha_2 p,\nonumber\\
\tilde{y}&=& \alpha_2 \rho + \alpha_3 p,
\end{eqnarray}
where 
\begin{eqnarray}
\alpha_1 = \frac{2+3\eta}{2+4\eta};\quad\alpha_2 =\frac{\eta}{2+4\eta},\quad\alpha_3 =\frac{2+\eta}{2+4\eta}.
\label{alphas}
\end{eqnarray}

Under the assumption that the temporal and spatial coefficients exhibit a Schwarzschild like solution, these definitions allow a GR like representation of TOV equations in this non-conservative gravity theory given as \cite{2015_Oliveira}
\begin{eqnarray}\label{eq9}
\frac{d\tilde{y}}{dr} = -\frac{\tilde{m} \tilde{x}}{r^2}\frac{\left(1+\frac{\tilde{y}}{\tilde{x}}\right)\left(1+\frac{4\pi r^3 \tilde{y}}{\tilde{m}}\right)}{1 - \frac{2\tilde{m}}{r}},
\end{eqnarray}
and 
\begin{eqnarray}\label{eq10}
\frac{d\tilde{m}}{dr}&=&4\pi \tilde{x} r^2.
\end{eqnarray}
In this parametric definition, $\lambda = 1+\eta$ and the new quantity $\eta$ parameterizes deviations from GR.

One may note that one can either solve the Eq.s \eqref{com00}, \eqref{com11} and \eqref{com22} numerically under the Schwarzschild like ansatz or can also solve the parameterized GR-like TOV equations to obtain the stellar structures after choosing a suitable EoS for the stellar configuration.

}

The value of $\eta$ can be weakly constrained by the energy conditions and stability of TOV equations. For the stability of TOV equations, one must have $\tilde{x}>0$ and $\tilde{y}>0$. These conditions in light of the MIT Bag model can be reduced to two scenarios {\it viz.,} $p(r)\leq 0$ and $p(r)> 0$. For the first scenario, one can have 
\begin{equation}
    B>-\frac{p(r)}{2}\land \left\lbrace\eta <\frac{-4 B-p(r)}{6 B+3 p(r)}\lor \eta >-\frac{p(r)}{2 B+p(r)}\right\rbrace. 
\end{equation}
While the second scenario returns three possibilities: 
\begin{equation}
\Big[0<B<\frac{p(r)}{2}\land \big\lbrace\eta <-\frac{p(r)}{2 B+p(r)}\lor \eta >\frac{-4 B-p(r)}{6 B+3 p(r)}\big\rbrace\Big]
\end{equation}
or, 
\begin{equation}
 \Big[B=\frac{p(r)}{2}\land \big\lbrace\eta <\frac{-4 B-p(r)}{6 B+3 p(r)}\lor \eta >\frac{-4 B-p(r)}{6 B+3 p(r)}\big \rbrace\Big]
\end{equation}
or, 
\begin{equation}
 \Big[B>\frac{p(r)}{2}\land \big\lbrace\eta <\frac{-4 B-p(r)}{6 B+3 p(r)}\lor \eta >-\frac{p(r)}{2 B+p(r)}\big \rbrace\Big]. 
\end{equation}

Here, one may note that choosing $p(r)\leq 0$ does not always create a stability issue due to the fact that in Rastall gravity effective pressure term is defined by $\tilde{y}$. Now, to have an approximate quantitative idea of the range of the parameter $\eta$, we consider the second scenario with two extreme cases, near the centre and near the surface of the compact star. Near the centre, $\tilde{y}$, as well as $\tilde{x}$, is expected to be larger than $B$. Hence, we have $\big\lbrace\eta <-\frac{p(r)}{2 B+p(r)}\lor \eta >\frac{-4 B-p(r)}{6 B+3 p(r)}\big\rbrace$, which can be approximately reduced to $\big\lbrace\eta <-1\lor \eta >\frac{-1}{3}\big\rbrace$. Near the surface of the star, $\tilde{y}$ is expected to be very small in comparison to $B$, but a finite density still contributes to $p$ to make it comparable to the order of $B$ ( see Eq. \ref{eq11}). Hence, at the surface, non-zero $p$ still allows us to choose $\eta<0$ in a stable hydrostatic equilibrium. However, $\eta<0$ might reduce the radii of the stars creating a more dense configuration. We shall investigate this effect extensively in the next part of our study.

\section{Equations of state}\label{sec03}
Solving the TOV equations can enable one to determine certain physical properties of compact stars, including their mass, radius, and more. However, this is only possible when these equations are accompanied by equations of state (EoSs) that establish relationships between energy density and fluid pressure. To describe dense matter in this study, we will use two EoSs. The first EoS we will utilize is the MIT Bag model EoS, which is the most commonly used and simplest EoS. This EoS, described by Witten in \cite{1984_Witten}, takes the form:

\begin{equation}
\label{n}
p=\frac{1}{3}(\rho-4\,B).
\end{equation}

It characterizes deconfined quark matter that is composed of $u$, $d$, and $s$ quarks. The confinement pressure is determined by the Bag constant $B$.
It is worth noting that the process of solving the TOV equations in conjunction with the MIT Bag model can yield stellar configurations that lack the necessary degree of compactness to feature a photon sphere around their surface, regardless of the various $B$ values within the accepted range \cite{jb1}. A photon sphere refers to the circular orbit of light that is a characteristic of highly compact objects such as black holes. However, prior research has indicated that this EoS can be stiffened by modification, as demonstrated by \cite{2018_manarelli, jb1, jb2, jb3}, with the following form:

\begin{equation}
\label{o}
p=\rho-4\,B.
\end{equation}

In this study, we will employ this modified and stiffer version of the MIT Bag model EoS. This EoS has been shown to be capable of generating the desired level of compactness, thus enabling the resulting stellar configurations to exhibit a photon sphere around their surface. A stiffer EoS refers to one that can support a higher pressure for a given energy density, which translates to a more compact and massive star. By using this modified version of the MIT Bag model EoS, we can explore the physical characteristics of compact stars more thoroughly and accurately. One may note that the maximum mass $M$ and corresponding radius $R$ of a compact star depends on the Bag constant $B$ in such a way that the compactification factor becomes independent of it. As $M\propto B^{-1/2}$ and $R\propto B^{-1/2}$ \cite{1984_Witten} respectively implying compactness $M/R$ is independent of $B$. In this regard, for the MIT Bag model we have chosen $B={(145\,\mbox{MeV})}^4$.

The second strange quark matter EoS that we will be examining is the CFL phase EoS \cite{alford}, as previously mentioned. The CFL phase involves the creation of $ud$, $us$, and $ds$ Cooper pairs, and its corresponding thermodynamic potential, $\Omega_{CFL}$, is of order $\Delta^2$. This potential can be obtained by using the following formula \cite{alford,lugones}:

\begin{equation}
\label{p}
\Omega_{CFL}=\Omega_{free}-\frac{3 \Delta^2 \mu^2}{\pi^2} +B,
\end{equation}
\begin{figure}
    \centering
      	\includegraphics[width=8.5cm,height=6cm]{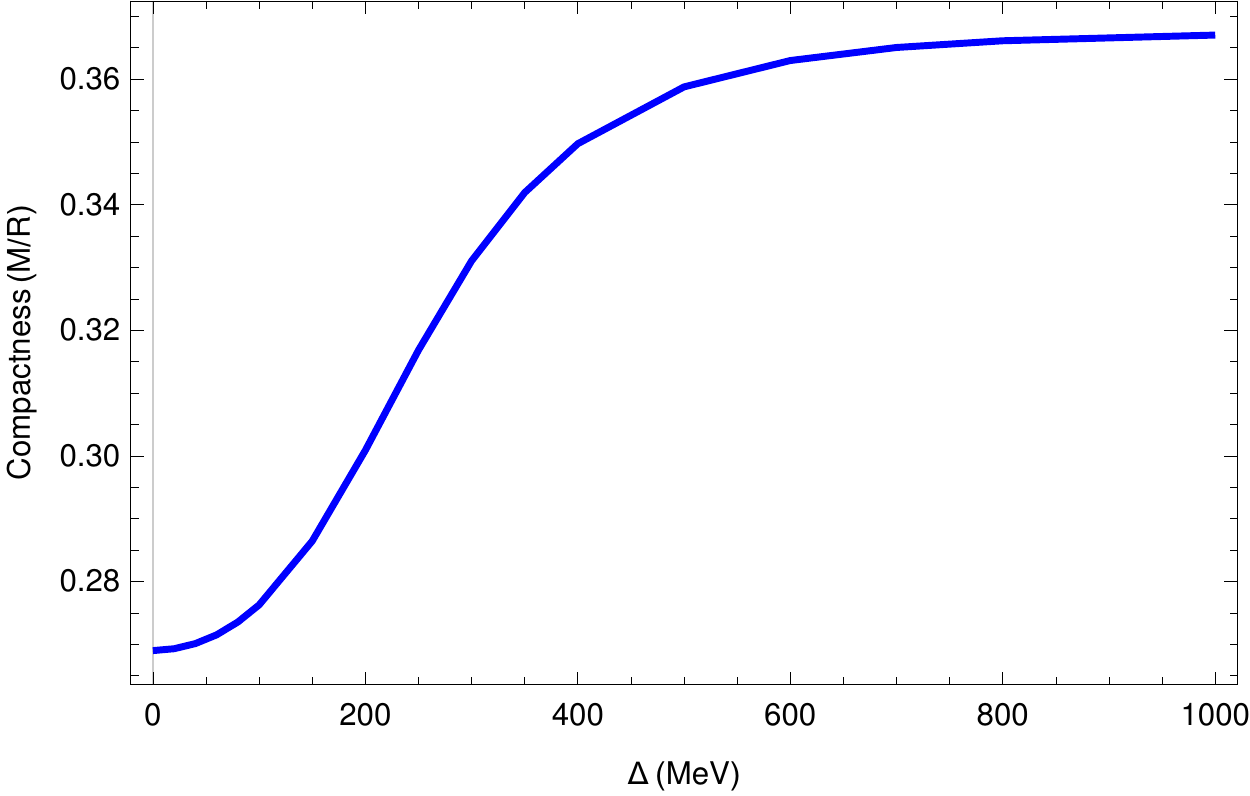} 
            	\caption{{Variation of compactness (M/R) with pairing gap $\Delta$ for the CFL phase of quark state. Here the value of Rastall parameter $\eta=-0.1$ is considered.}}\label{fig00}
\end{figure}

where $\Omega_{free}$ represents the thermodynamic potential of the free quarks without pairing interaction. {$\Delta$ is the pairing energy gap, $\mu$ represents the baryon chemical potential and $B$ is the Bag constant as defined earlier.} It is important to note that the CFL phase EoS offers a unique perspective on the properties of strange quark matter due to its unconventional pairing mechanism. By exploring this EoS, we can gain a deeper understanding of the physical behavior of compact stars and their constituents. The expression for $\Omega_{free}$ is given by
\begin{small}   
\begin{eqnarray}
\label{q}
\Omega_{free}=\frac{6}{\pi^2}\int_{0}^{\nu}\big[p-\mu\big]p^2dp+\frac{3}{\pi^2}\int_{0}^{\nu}\left[(p^2+m_s^2)^{1/2}-\mu\right]p^2dp,\\
=\sum_{i\,=\,u,d,s}\frac{1}{4\pi^2}\left[\mu_i \nu\left(\mu_i^2-\frac{5}{2}m_i^2\right)+\frac{3}{2}m_i^4 \log\left(\frac{\mu_i+\nu}{m_i}\right)\right],
\end{eqnarray}
\end{small}
where $m_s$ is the mass of strange quark, $\mu$ is the baryon chemical potential and $3\mu=\mu_u+\mu_d+\mu_s$.

As a result of the pairing interaction forces, the different flavors of quarks possess the same baryon number density $n_B$ and particle number densities, as shown below:
\begin{equation}
\label{r}
n_B=n_u=n_d=n_s=\frac{(\nu^3+2\Delta^2\mu)}{\pi^2}
\end{equation}

The common Fermi momentum is represented by $\nu$, which is given by the square root of the difference between the square of the chemical potential and the square of the mass. Specifically, $$\nu=2\mu-\left(\mu^2+\frac{m_s^2}{3}\right)^{1/2}$$, where $\mu_i$ and $m_i$ are the chemical potential and mass of the $i$-th quark flavor, respectively. The pairing gap, denoted by $\Delta$, can be considered as a free parameter \cite{lugones} and is related to the QCD Cooper pairs. The corresponding condensate term is given by $3 \Delta^2 \mu^2/\pi^2$. The pressure and energy density of strange quark matter (SQM) can be calculated from the thermodynamic potential as shown below:
\begin{equation}
\label{s}
p=-\,\Omega_{CFL},
\end{equation}
\begin{equation}
\label{t}
\rho=\sum_i \mu_i n_i+\Omega_{CFL}=3\,\mu\, n_B-p.
\end{equation}

This EoS for quark matter based on the CFL phase state can now be expressed as
\begin{equation}\label{u}
\rho=3\,p+4B-\frac{9\,\xi\,\mu^2}{\pi^2},
\end{equation}
or
\begin{equation}
\label{v}
p=\frac{\rho}{3}-\frac{4B}{3}+\frac{3\,\xi\,\mu^2}{\pi^2},
\end{equation}
here $\mu^2$ and $\xi$ are given by
\begin{eqnarray}
\label{w}
\mu^2&=&-\,3\,\xi+\left[9\,\xi^2+\frac{4}{3}\pi^2(p+B)\right]^{1/2} \nonumber \\
&=&-\,\xi+\left[\xi^2+\frac{4}{9}\pi^2(\rho-B)\right]^{1/2}\!\!\!,
\end{eqnarray}
and \begin{equation}
\label{y}
\xi=-\frac{m_s^2}{6}+\frac{2\Delta^2}{3}.
\end{equation}

We can regard the parameters $\Delta$ and $m_s$ as free parameters and confine them using stability criteria as we lack the appropriate values \cite{jb4,2019_Maulana,flores,farhi}. In the present work, we have considered a finite mass case $m_s\neq0$ considering $m_s=100$ MeV \cite{beringer}. {Also we have chosen $B = {(180\,\mbox{MeV})}^4$ for this EoS. It is to be noted here that, with these $B$ and $m_S$ values, we have considered a high value of the pairing constant $\Delta$. The reason for choosing high $\Delta$ is to obtain stars having compactness $\geq 0.33$. In this regard, the variation of compactness of strange stars with different $\Delta$ values can be useful. In Fig. \ref{fig00}, such a variation for Rastall gravity with $\eta=-0.1$ is shown. It is noticed that for $\Delta\approx300$ MeV desired compactness is approached. However, for absolutely bound CFL state one can consider the stability window for the CFL phase state (the $m_s-B$ plane) as discussed in the Ref. \cite{lugones}. In the present article, regardless of the absolute stability we are interested in considering the echo possibilities from these stars \cite{jb4}.}

{ \section{Energy conditions}
In the following section, an examination of the energy conditions will be undertaken, relying on the empirical findings of Hanafy et al.~\cite{hanafy}.  In the framework of General Relativity (GR), it was established the positive nature of both the trace of the tidal tensor $R_{\epsilon\upsilon} U^{\epsilon} U^{\upsilon} $ and the $R_{\epsilon\upsilon} l^{\epsilon} l^{\upsilon} $ term in the Raychaudhuri equation, where $U^{\epsilon}$  denotes the arbitrary time-like vector while $l^{\epsilon}$ is an arbitrary null vector going towards the future. There are four constraints that are imposed on the EMT ($T_{\epsilon \upsilon}$), which can be referred to as the energy conditions. These concepts have the potential to be used to the field of modified gravity. In the specific instance of Rastall's theory of gravity, the energy conditions may be expressed in relation to the effective energy-momentum tensor. So, first we need to define the effective energy momentum tensor. We have from the Eq.(\ref{eq8}) and Eq.(\ref{eq9}), 
\begin{eqnarray}
 G_{\epsilon\upsilon} +\gamma g_{\epsilon\upsilon} R &=& k T_{\epsilon\upsilon}, \label{eqE11} 
\end{eqnarray}
After contracting Eq. (\ref{eqE11}),  we may rewrite it as follows: 
\begin{eqnarray}
 R=-\frac{k}{1-4\gamma} T, ~~~~~~\left(\gamma\ne\frac{1}{4}\right)\label{eqE12} 
\end{eqnarray}
Here, $T=g^{\epsilon\upsilon}T_{\epsilon\upsilon}$ represents the trace of the energy-momentum tensor. Using Eq.(\ref{eqE12}), the field equation (\ref{eqE11}) may be expressed as
\begin{eqnarray}
 G_{\epsilon\upsilon} &=& k\, T^{\text{eff}}, \label{eqE13} 
\end{eqnarray}
where, 
\begin{eqnarray}
    T^{\text{eff}}=\left(T_{\epsilon\upsilon}+\frac{\gamma}{1-4\gamma} g_{\epsilon\upsilon} T \right); ~~~~~~\left(\gamma\ne\frac{1}{4}\right).
\end{eqnarray}
In order for a physical model to be considered valid, it must adhere to the modified energy conditions outlined below: 
\begin{itemize}
    \item for weak energy condition (WEC): $\rho^{\text{eff}} \ge 0$ and\\ $\rho^{\text{eff}}+p^{\text{eff}} \ge 0$;
    \item for null energy condition (NEC): $\rho^{\text{eff}}+p^{\text{eff}} \ge 0$;
   \item  for strong energy condition (SEC): $\rho^{\text{eff}}+p^{\text{eff}} \ge 0$, and $\rho^{\text{eff}}+3p^{\text{eff}} \ge 0$;
   \item for dominant energy conditions (DEC): $\rho^{\text{eff}}-p^{\text{eff}} \ge 0$. 
\end{itemize}
where,
\begin{eqnarray}
   && \rho^{\text{eff}}= \rho+\frac{\gamma}{1-4\gamma} (\rho-3p),\\
   && p^{\text{eff}}= p- \frac{\gamma}{1-4\gamma} (\rho-3p).
\end{eqnarray}

since, $\gamma =(1-\lambda)/2$ and $\lambda=1+\eta$, then we get the following relation between $\gamma$ and $\eta$ as: $\gamma=-\eta/2$. Then the effective pressure and effective density can be written in term of parameter $\eta$ as:  \begin{eqnarray}
   && \rho^{\text{eff}}= \rho-\frac{\eta}{2 + 4 \eta} \left(\rho-3p\right),\\
   && p^{\text{eff}}= p+\frac{\eta}{2 + 4 \eta} \left(\rho-3p\right).
\end{eqnarray}}

\begin{figure*}
      	\centering
     \includegraphics[width=5.5cm,height=5.5cm]{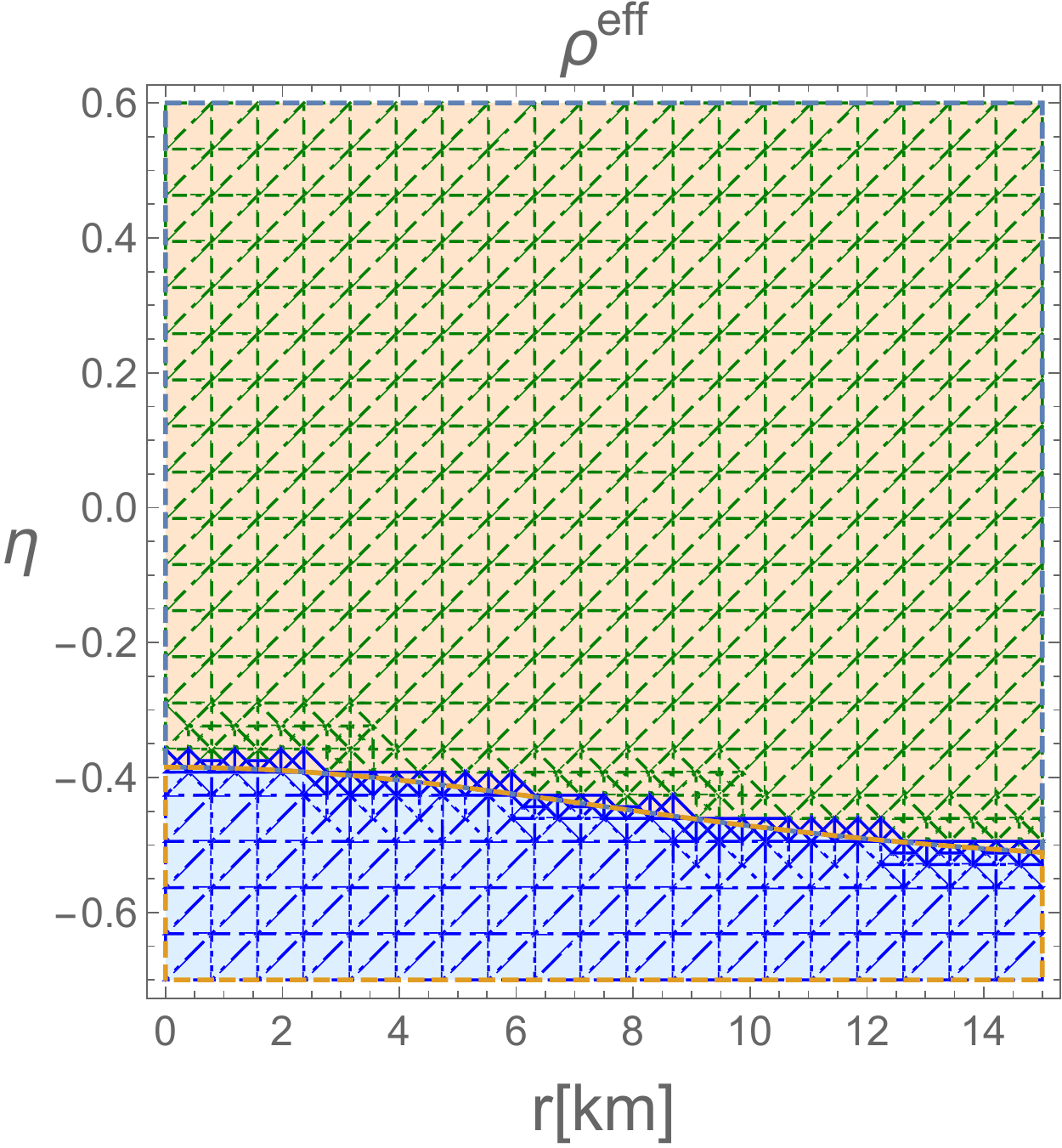}
      \includegraphics[width=5.5cm,height=5.5cm]{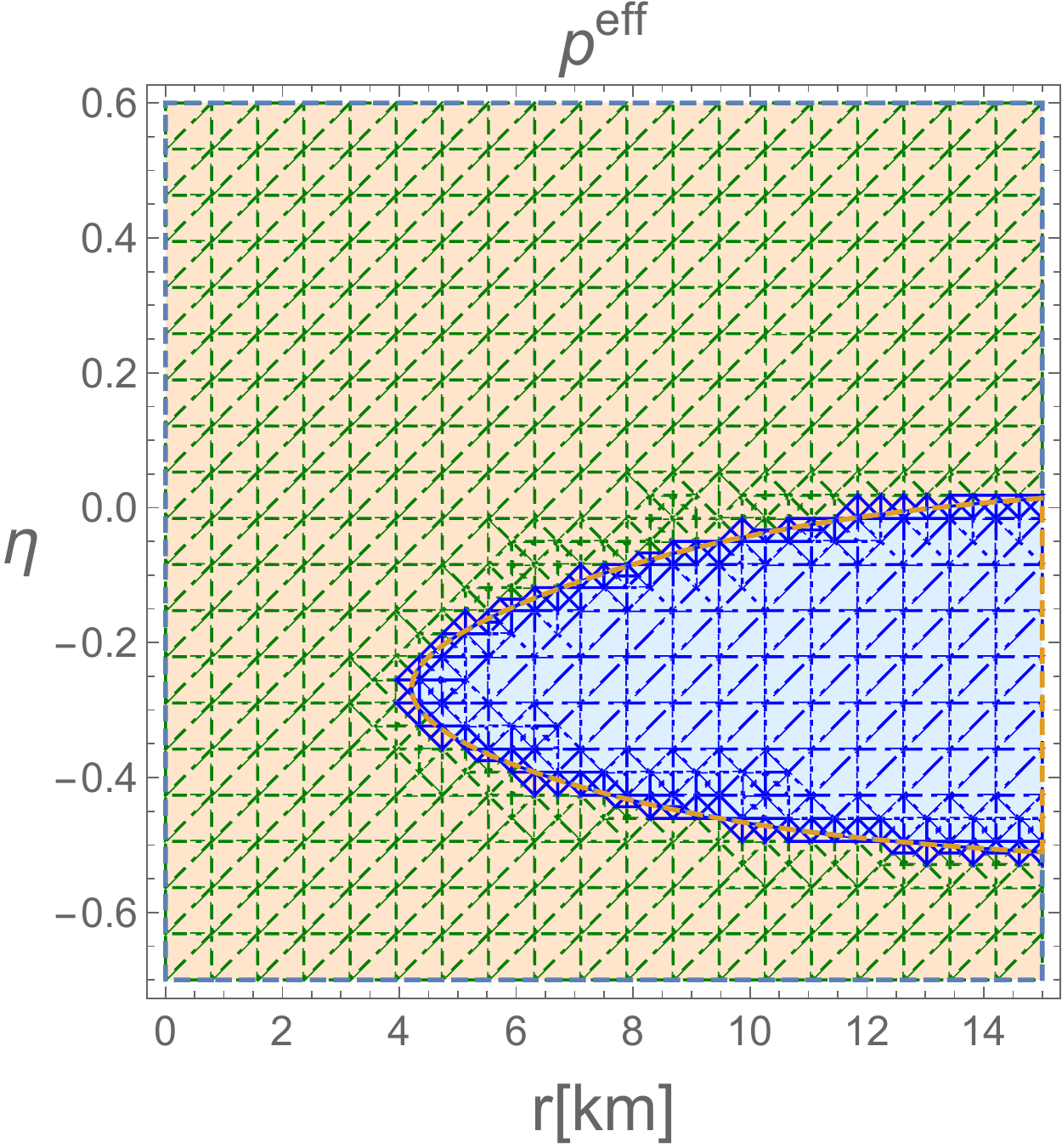} 
      \includegraphics[width=5.5cm,height=5.5cm]{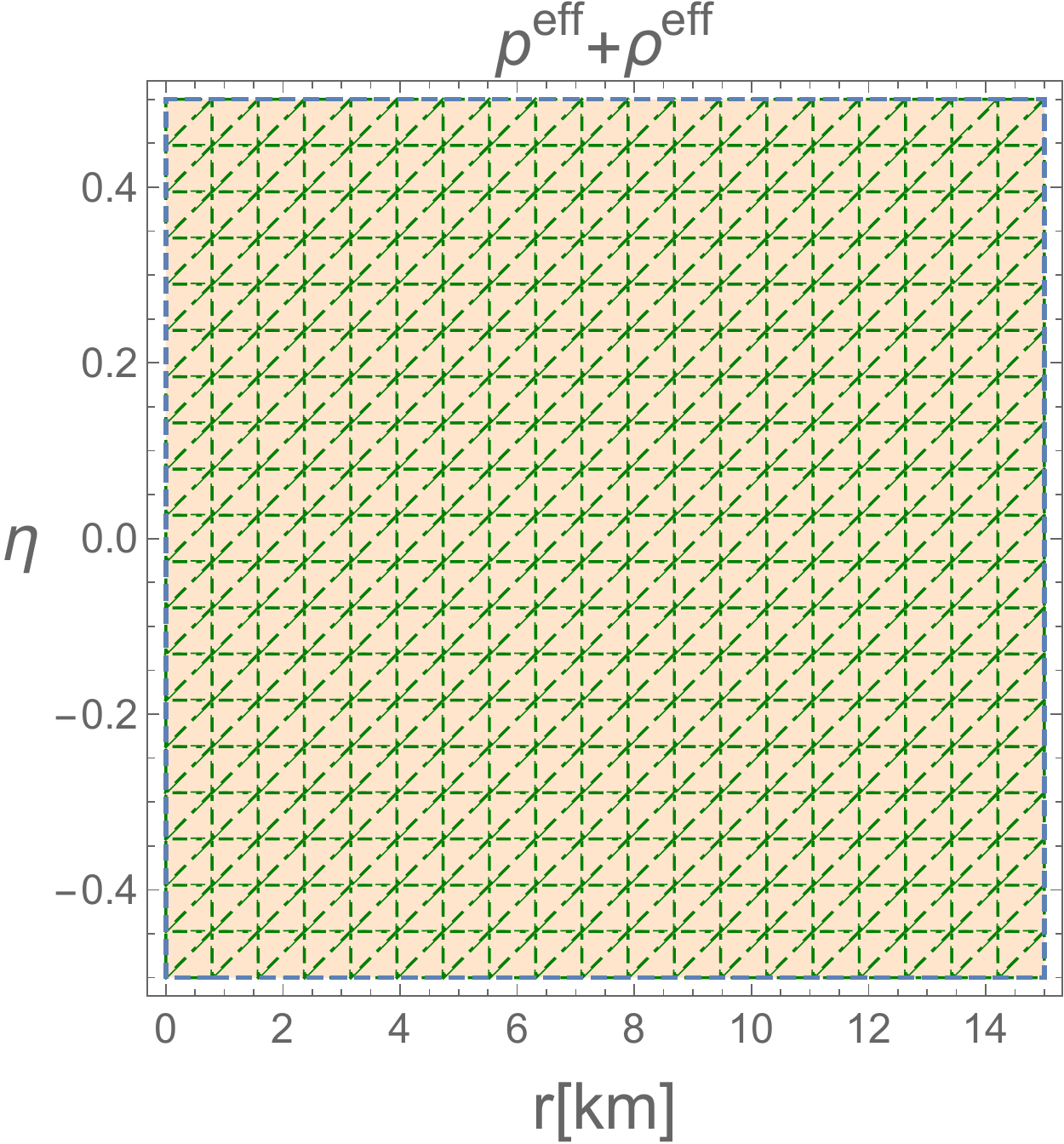} 
       \includegraphics[width=5.5cm,height=5.5cm]{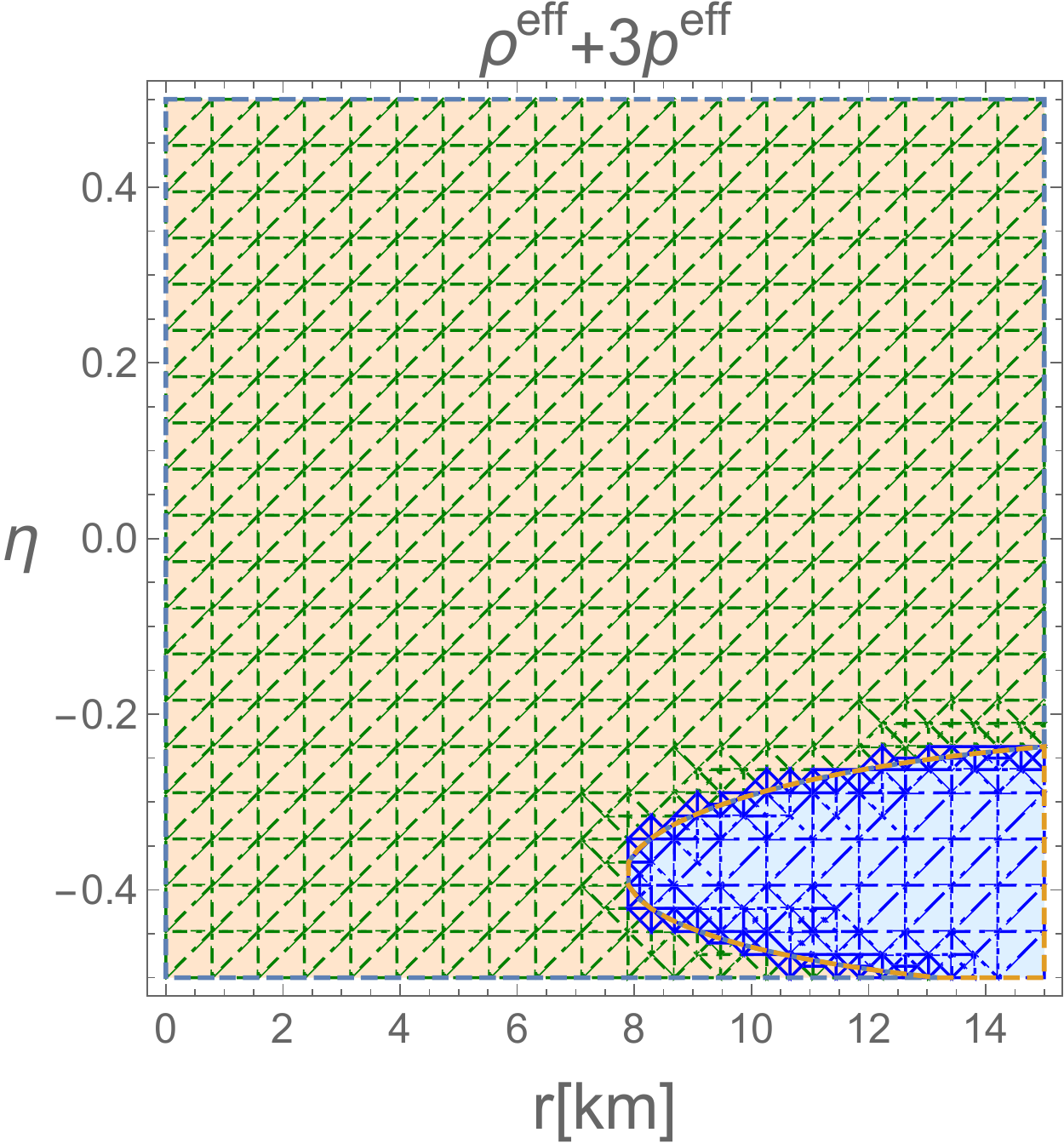}
      \includegraphics[width=5.5cm,height=5.5cm]{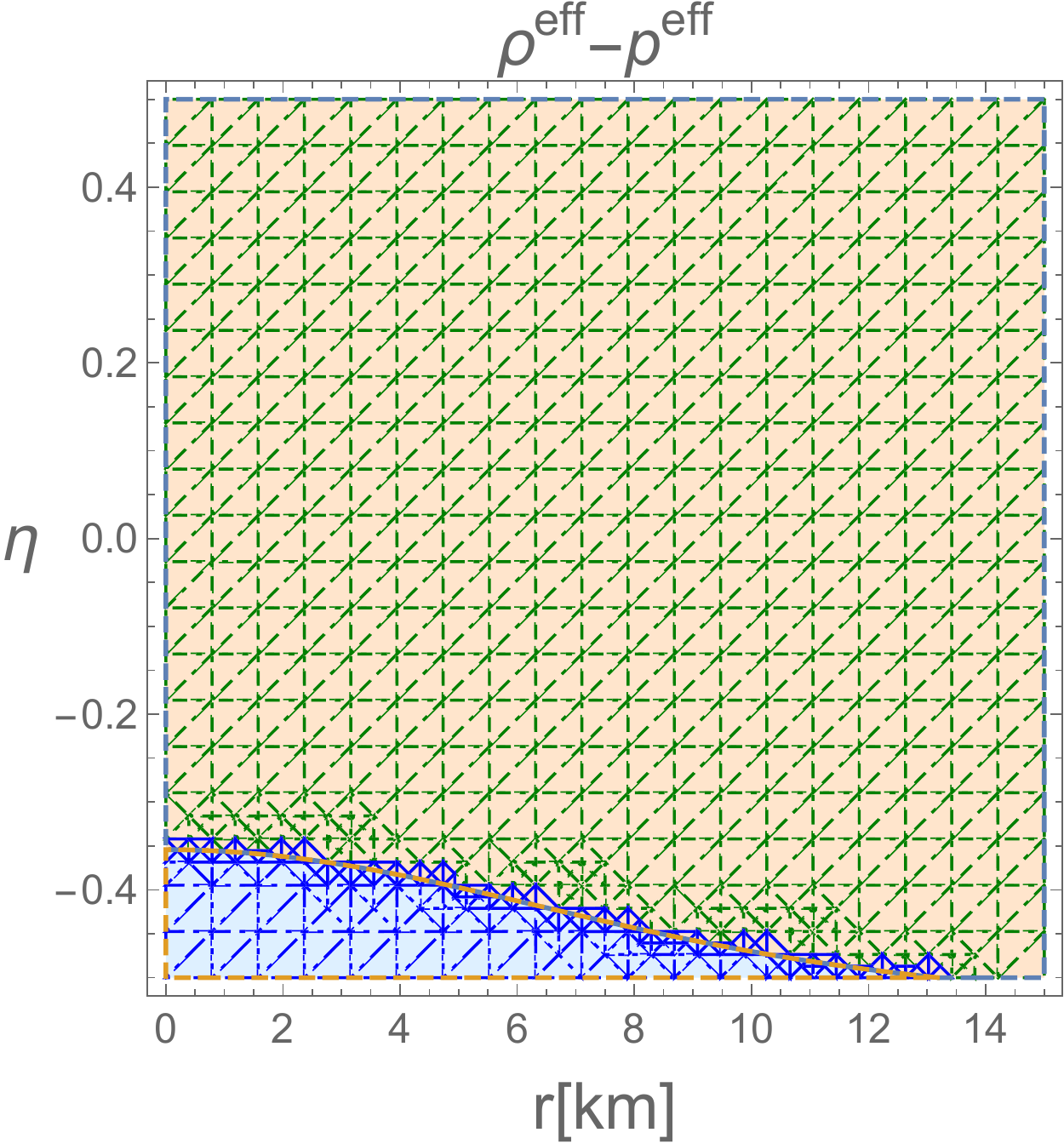}
      \includegraphics[width=5.5cm,height=5.5cm]{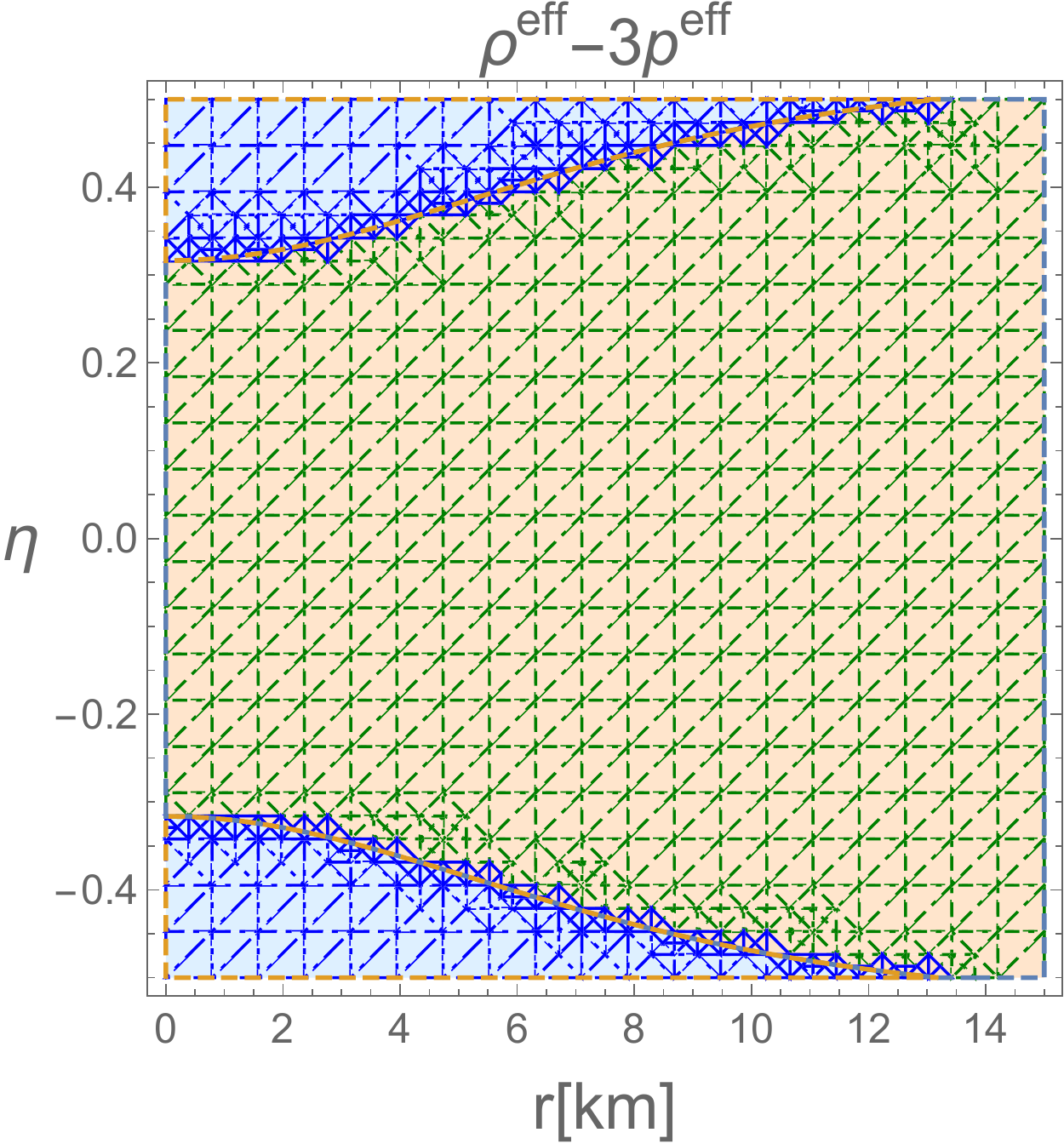}
            	\caption{Region plots of $\rho ^{\text{eff}}$, $p^{\text{eff}} $, $\rho ^{\text{eff}}+p^{\text{eff}}$, $\rho ^{\text{eff}}+3p^{\text{eff}}$, $\rho ^{\text{eff}}-p^{\text{eff}}$, and $\rho ^{\text{eff}}-3p^{\text{eff}}$ (the feasible region is depicted by green shaded region and unfeasible region is depicted by blue shaded region) of compact stars with stiffer MIT model for $B = {(145\,\mbox{MeV})}^4$.}\label{fig0}
      \end{figure*}

{By utilizing the EoS (\ref{o}) and choosing the central density around $10B$, we have shown the viable regions of the parameter space in Fig. \ref{fig0}. One can see that $\rho^{\text{eff}} + p^{\text{eff}}>0$ is satisfied up to approximately $r=15$ km for both positive and negative values of the Rastall parameter. Since, physical compact star candidates do not have a very large radius, $\rho^{\text{eff}} + p^{\text{eff}}>0$ is satisfied for both positive and negative values of the Rastall parameter $\eta$.  For $\rho^{\text{eff}}>0$, we have noticed a decreased feasibility region in the $r-\eta $ parameter space. Again, in the case of $p^{\text{eff}}>0$, the feasibility region is found to decrease near the larger radii regions. 
Similarly, we have plotted the region plots for $\rho^{\text{eff}}+3p^{\text{eff}}$, $\rho^{\text{eff}} - p^{\text{eff}}$, and $\rho^{\text{eff}}-3p^{\text{eff}}$ to see the feasible parameter spaces allowed by the energy conditions. In our analysis, we choose the Rastall parameter values from these feasible parameter spaces to avoid instabilities.}

\section{Physical properties}\label{sec05}
\subsection{Mass-radius profiles}
\begin{figure}
      	\centering{
      	\includegraphics[width=10cm,height=8cm]{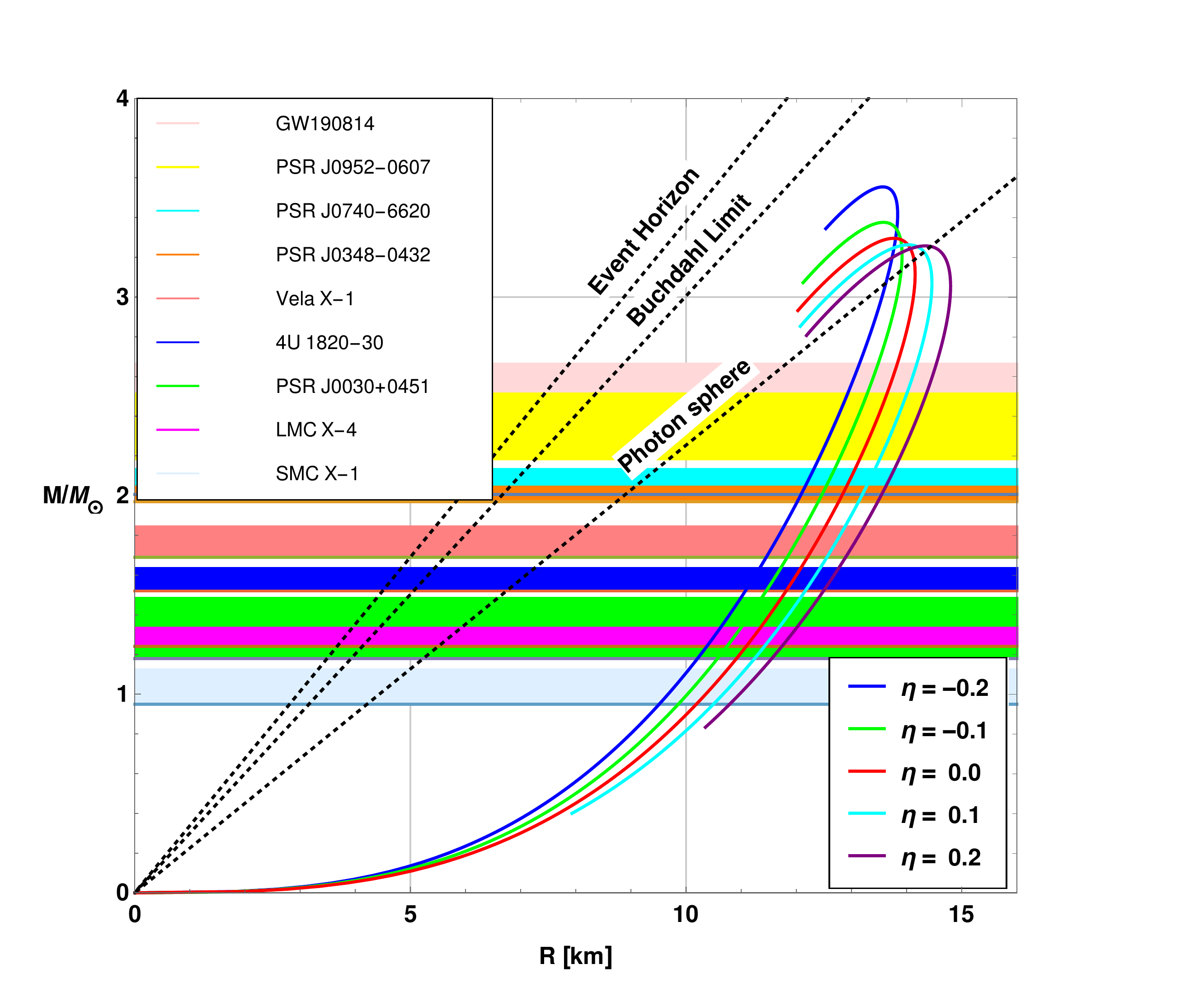}}
      	\caption{Variation of mass with respect to the radius of compact stars with stiffer MIT model for $B = {(145\,\mbox{MeV})}^4$ and varying Rastall parameter $\eta$ along with constraints from astrophysical observations.}\label{fig1}
      \end{figure}

\begin{figure}
      	\centering{
      	\includegraphics[width=10cm,height=8cm]{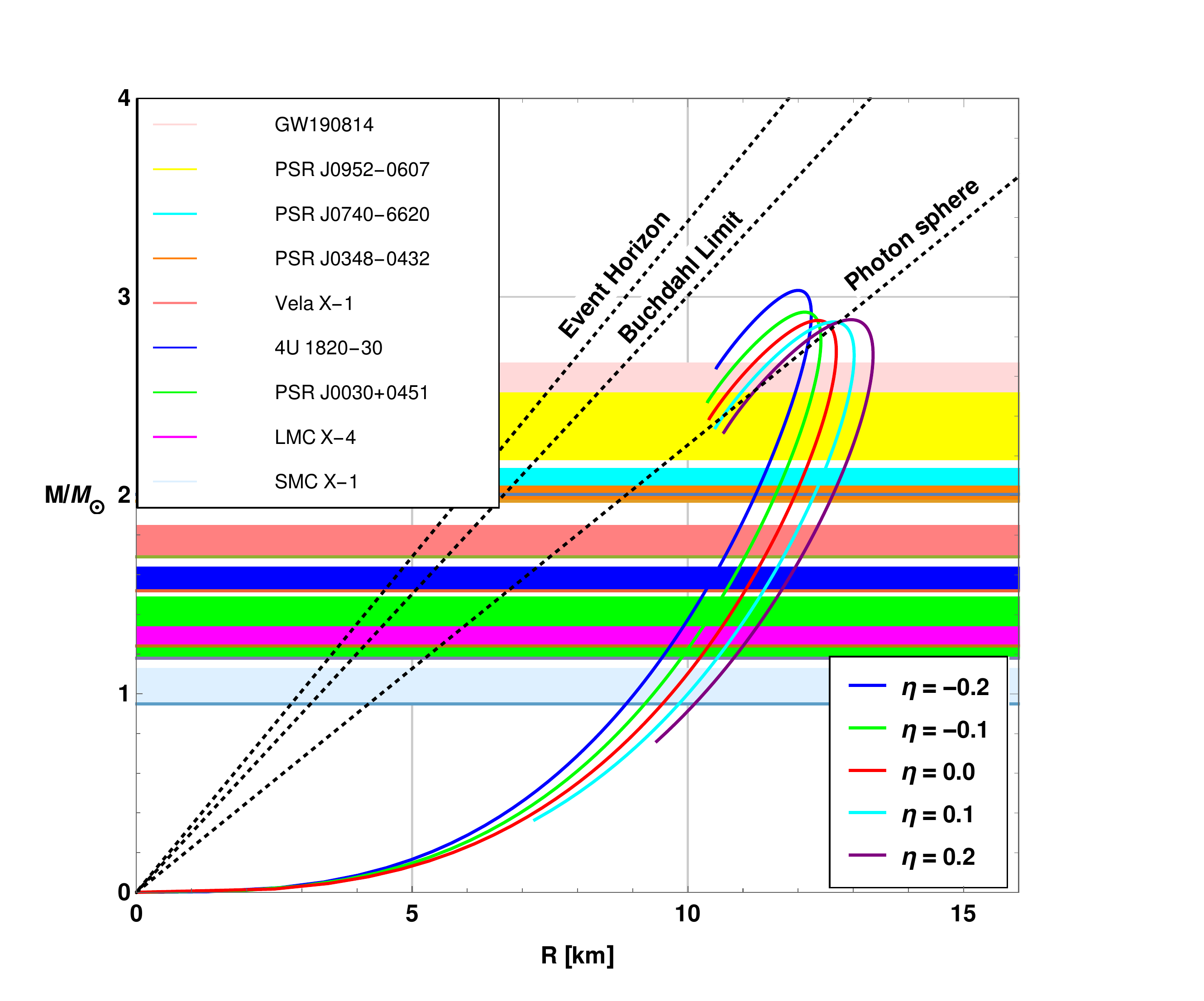}}
      	\caption{Mass-radius profile for compact stars with CFL Phase state with $m_s=100 MeV$, $B = {(180\,\mbox{MeV})}^4$ and $\Delta = 470 MeV$ and varying Rastall parameter $\eta$ along with constraints from astrophysical observations.}\label{fig2}
      \end{figure}

To collect knowledge on the physical properties of star models, an important point is to study the behaviour of mass with radius. From such relations, we can directly compare the values from observational results and that obtained from calculations. Thus, this relation is helpful while studying the EoSs of compact objects. For the considered cases of this study, we have drawn the mass-radius variation patterns. To draw these graphs, we have numerically solved the TOV equations in the Rastall gravity with suitable boundary conditions. {One may note that we have used the condition $p(R)=0$ to determine the radius of the strange star.} 
In Fig. \ref{fig1}, \ref{fig2}, the impacts on the mass-radius relations of strange stars are shown for different non-conservative terms $\eta$ of Rastall gravity. {In these figures, we have plotted the mass $M$ of strange stars with respect to the radial distance $R$.} 
In Fig. \ref{fig1}, the results are shown using the MIT Bag model EoS, and in Fig. \ref{fig2}, the relations obtained using the CFL phase state are shown. For these two EoSs of the strange star, it can be understood that for a fixed value of the Bag constant $B$ and strange-quark mass $m_s$, the mass-radius curves change significantly. More negative $\eta$ values correspond to stars with a more compactification factor and vice-versa for both EoSs. 

      
\begin{table*}
\centering
\caption{\label{table1} Predicted radius of strange star candidates with MIT EoS for different $\eta$ values.}
\scalebox{1}{\begin{tabular}{|*{7}{c|}}
\hline
{Candidates} & {Mass M (in $\mbox{M}_{\odot}$)}  & \multicolumn{5}{c|}{{Predicted Radius $R$ (in $\mbox{km}$)}} \\
\cline{3-7}
&& \multicolumn{5}{c|}{$\eta$} \\
\cline{3-7}
&  & -0.2 & -0.1 & 0 & 0.1 & 0.2   \\ \hline
GW190814 \cite{gw190814}  &  2.5 - 2.67  & 12.91 - 13.51  &  13.26 - 13.48  & 13.62 - 13.83 & 13.98 - 14.19 & 14.34 - 14.54 \\
PSR J0952-0607 \cite{romani} &  2.35 $\pm$ 0.17  & $12.67^{+0.26}_{-0.28}$  & $13.04^{+0.24}_{-0.27}$ & $13.41^{+0.23}_{-0.27}$ & $13.77^{+0.24}_{-0.27}$ & $14.12^{+0.23}_{-0.27}$ \\
PSR J0740+6620 \cite{6620_1,6620_2} & $2.072^{+0.067}_{-0.066}$ & $12.20^{+0.119}_{-0.128}$  & $12.58^{+0.11}_{-0.12}$ & $12.95^{+0.11}_{-0.12}$ & $13.31^{+0.11}_{-0.12}$ & $13.66^{+0.11}_{-0.12}$  \\
PSR J0348+0432 \cite{antoni} & 2.01 $\pm$ 0.04 & $12.09^{+0.073}_{-0.074}$  & $12.47^{+0.072}_{-0.074}$ & $12.84^{+0.06}_{-0.08}$ & $13.20^{+0.07}_{-0.08}$ & $13.55^{+0.06}_{-0.12}$  \\
Vela X-1 \cite{gango, rawls} & 1.77 $\pm$ 0.08 & $11.62^{+0.162}_{-0.168}$ & $12.00^{+0.162}_{-0.169}$ & $12.37^{+0.16}_{-0.17}$ & $12.72^{+0.16}_{-0.17} $ & $13.07^{+0.15}_{-0.18}$ \\
4U 1820-30 \cite{guver} & 1.58 $\pm$ 0.06  & $11.21^{+0.133}_{-0.137}$ & $11.59^{+0.134}_{-0.139}$ & 11.95 $\pm$ 0.14 & $12.30^{+0.13}_{-0.14}$ & $12.63^{+0.13}_{-0.14}$ \\
PSR J0030+0451 \cite{0451,0451_2} &$1.34^{+0.15}_{-0.16}$& $10.63^{+0.36}_{-0.42}$ & $11.00^{+0.37}_{-0.43}$ & $11.36^{+0.38}_{-0.45}$ & $11.69^{+0.38}_{-0.45}$ & $12.02^{+0.38}_{-0.47}$ \\
LMC X-4	\cite{rawls} & 1.29 $\pm$ 0.05  & $10.50^{+0.129}_{-0.133}$ & $10.87^{+0.131}_{-0.135}$ & 11.22 $\pm$ 0.14 & $11.56^{+0.12}_{-0.15}$ & 11.87 $\pm$ 0.14  \\
SMC X-1	\cite{aziz} & 1.04 $\pm$ 0.09  & $9.79^{+0.26}_{-0.28}$ & $10.15^{+0.27}_{-0.29}$ & $10.48^{+0.28}_{-0.30}$ & $10.80^{+0.28}_{-0.31}$ & $11.10^{+0.28}_{-0.31}$\\
\hline
\end{tabular}}
\end{table*}

\begin{table*}
\centering
\caption{\label{table2} Physical properties of strange star candidates with CFL phase.}
\scalebox{1}{\begin{tabular}{|*{7}{c|}}
\hline
{Candidates} & {Mass M (in $\mbox{M}_{\odot}$)}  & \multicolumn{5}{c|}{{Predicted Radius $R$ (in $\mbox{km}$)}} \\
\cline{3-7}
&& \multicolumn{5}{c|}{$\eta$} \\
\cline{3-7}
&  & -0.2 & -0.1 & 0 & 0.1 & 0.2   \\ \hline
GW190814 \cite{gw190814}  &  2.5 - 2.67  & 11.92 - 12.10  & 12.24 - 12.37 & 12.57 - 12.68 & 12.91 - 13.01 & 13.26 - 13.35  \\
PSR J0952-0607 \cite{romani} &  2.35 $\pm$ 0.17  & $11.74^{+0.20}_{-0.23}$ & $12.08^{+0.18}_{-0.22}$ & $12.42^{+0.16}_{-0.21}$ & $12.77^{0.16}_{-21}$ & $13.11^{+0.16}_{-0.21}$ \\
PSR J0740+6620 \cite{6620_1,6620_2} & $2.072^{+0.067}_{-0.066}$ & $11.33^{+0.103}_{-0.108}$ & $11.69^{+0.100}_{-0.105}$ & $12.05^{+0.09}_{-0.10}$ & $12.40^{+0.09}_{-0.10}$ & $12.74^{+0.100}_{-0.106}$ \\
PSR J0348+0432 \cite{antoni} & 2.01 $\pm$ 0.04 & $11.23^{+0.064}_{-0.066}$ & $11.60^{+0.06}_{-0.56}$ & $11.95^{+0.062}_{-0.064}$ & $12.30^{+0.062}_{-0.064}$ & $12.64^{+0.063}_{-0.065}$ \\
Vela X-1 \cite{gango, rawls} & 1.77 $\pm$ 0.08 & $10.81^{+0.14}_{-0.15}$ & $11.18^{+0.14}_{-0.15}$ & $11.54^{+0.14}_{-0.15}$ & $11.88^{+0.14}_{-0.15}$ & $12.21^{+0.150}_{-0.158}$ \\
4U 1820-30 \cite{guver} & 1.58 $\pm$ 0.06 & $10.44^{+0.121}_{-0.125}$ & $10.81^{+0.122}_{-0.126}$ & $11.16^{+0.123}_{-0.128}$ & $11.50^{+0.12}_{-0.13}$ & $11.82^{+0.12}_{-0.13}$ \\
PSR J0030+0451 \cite{0451,0451_2} & $1.34^{+0.15}_{-0.16}$ & $9.92^{+0.33}_{-0.39}$ & $10.28^{+0.34}_{-0.40}$ & $10.62^{+0.34}_{-0.40}$ & $10.95^{+0.35}_{-0.41}$ & $11.26^{+0.36}_{-0.42}$ \\
LMC X-4	\cite{rawls} & 1.29 $\pm$ 0.05 & $9.80^{+0.11}_{-0.12}$ & $10.16^{+0.121}_{-0.125}$ & $10.50^{+0.123}_{-0.127}$ & $10.82^{+0.12}_{-0.13}$ & $11.13^{+0.129}_{-0.133}$ \\
SMC X-1	\cite{aziz} & 1.04 $\pm$ 0.09 & $9.14^{+0.24}_{-0.26}$ & $9.49^{+0.25}_{-0.27}$ & $9.81^{+0.25}_{-0.27}$ & $10.12^{+0.26}_{-0.28}$ & $10.41^{+0.27}_{-0.29}$ \\
\hline
\end{tabular}}
\end{table*}
\begin{figure*}
      	\centering{
      	\includegraphics[width=8cm,height=6.5cm]{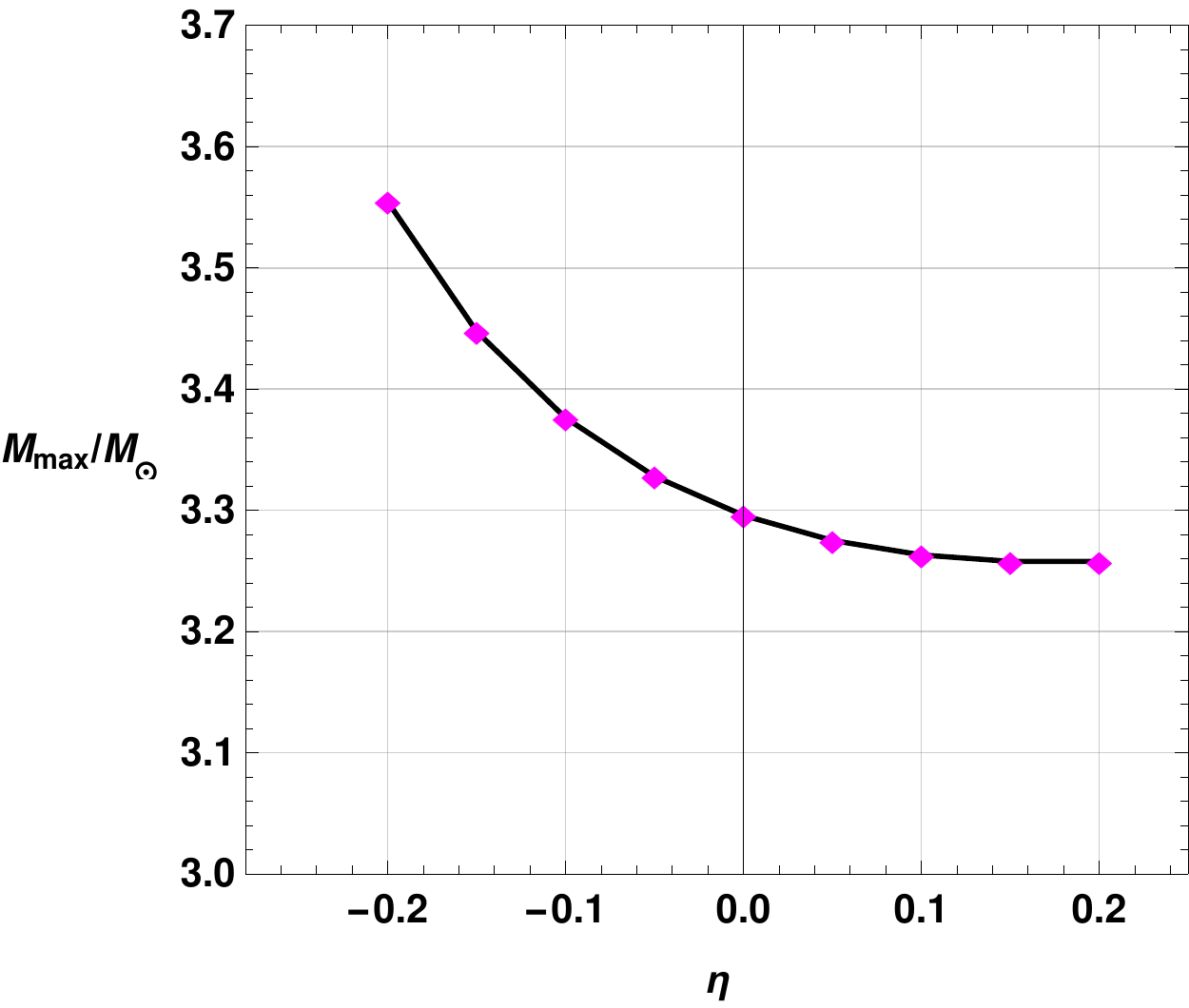}~~~~\includegraphics[width=8cm,height=6.5cm]{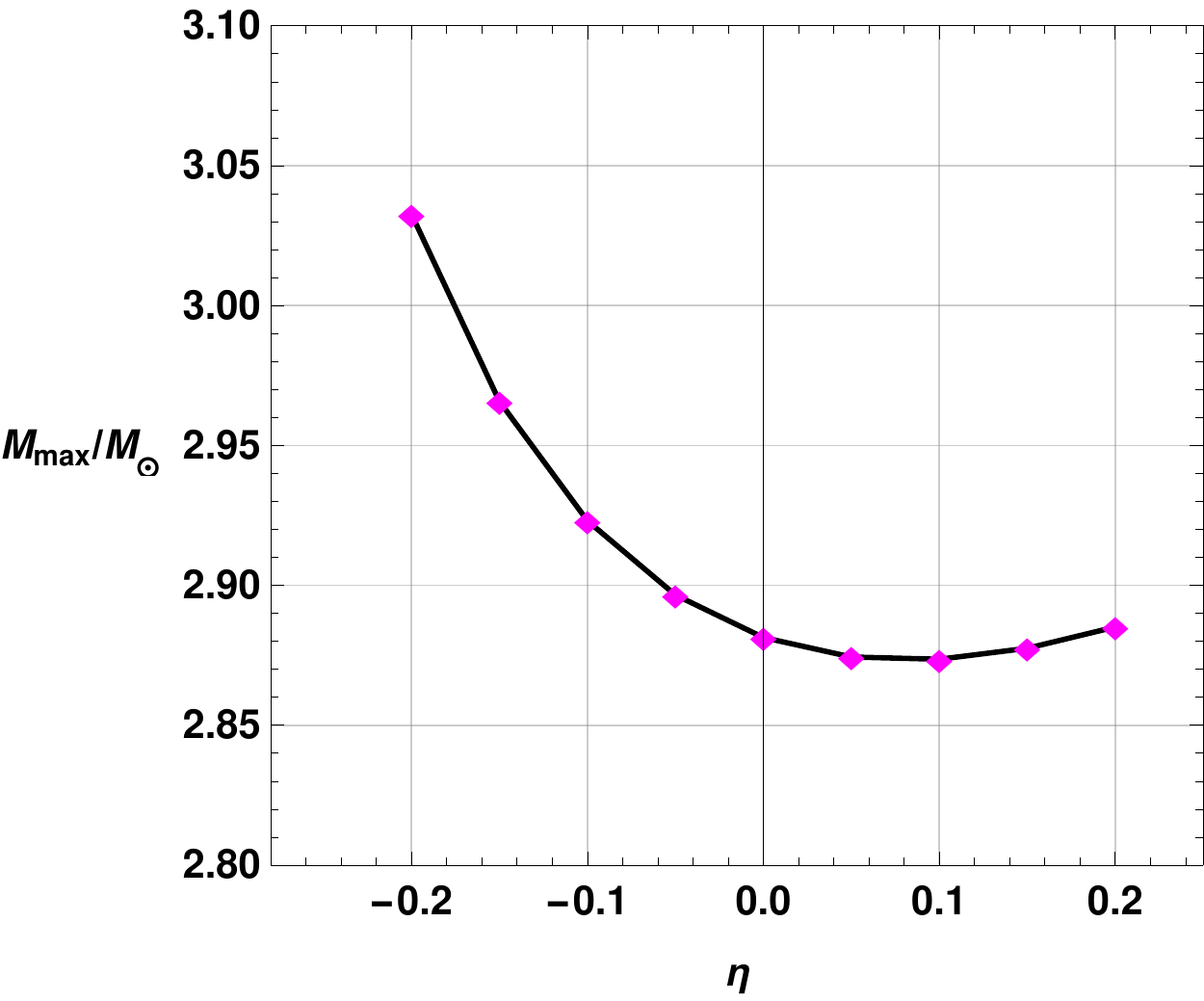}}
      	\caption{Variation of maximum mass with respect to Rastall parameter $\eta$ of compact stars with MIT Bag model having $B = {(145\,\mbox{MeV})}^4$ (top panel) and CFL Phase state with $m_s=100 MeV$, $B = {(180\,\mbox{MeV})}^4$ and $\Delta = 470 MeV$ (bottom panel).}
      	\label{fig3}
      \end{figure*}

\begin{figure*}
      	\centering{
      	\includegraphics[width=7.5cm,height=6.5cm]{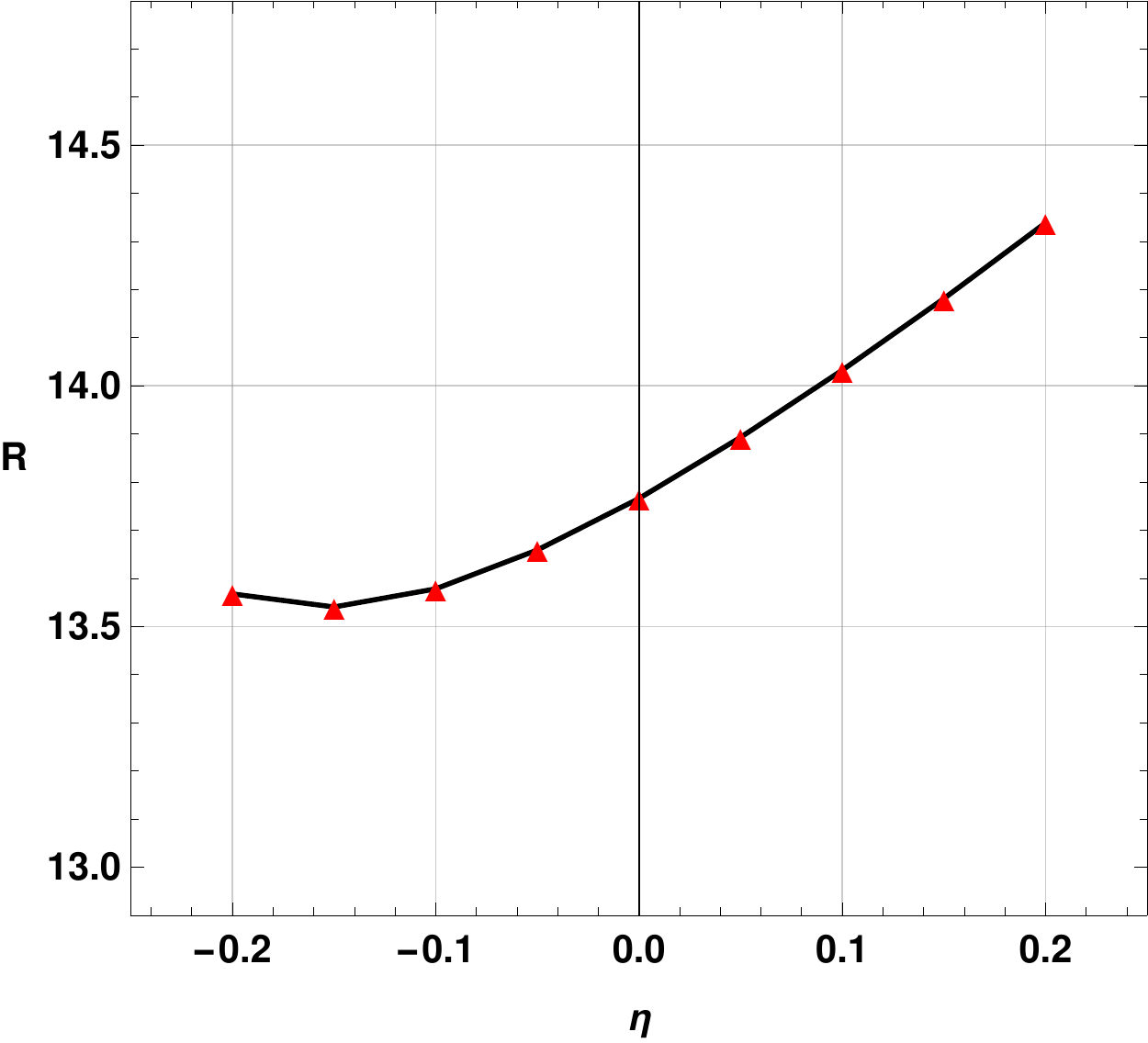}~~~~
            \includegraphics[width=7.5cm,height=6.5cm]{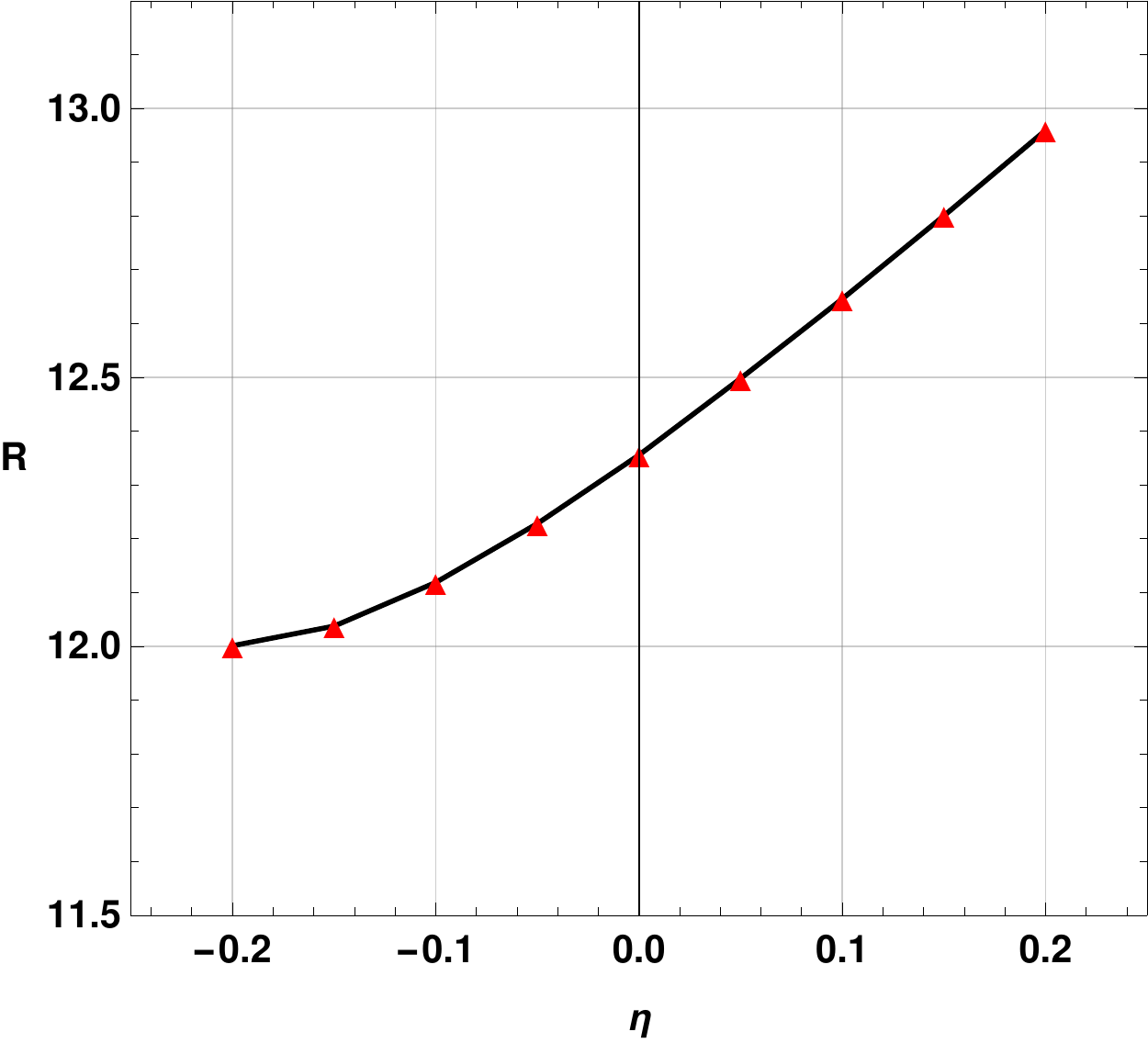}}
      	\caption{Variation of radius $R$ corresponding to maximum mass $M_{max}$ with respect to Rastall parameter $\eta$ of compact stars with MIT Bag model having $B = {(145\,\mbox{MeV})}^4$ (top panel) and CFL Phase state with $m_s=100 MeV$, $B = {(180\,\mbox{MeV})}^4$ and $\Delta = 470 MeV$ (bottom panel).}
      	\label{fig4}
      \end{figure*}

We represent these mass-radius relations along with mass-radius constraints obtained through a few important astrophysical observations. In light of these astrophysical observations discussed in Section \ref{sec01}, we have prepared a data sheet in Table \ref{table1} and \ref{table2} with their observational mass together with the predicted radius for different model parameters. For this, we have chosen the results for GW 190814 \cite{gw190814}, PSR J0030+0451 and PSR J0740+6620 \cite{0451,0451_2,6620_1,6620_2}, PSR J0952-0607 \cite{romani}, PSR J0348+0432 \cite{Tangphati1}. Besides the inclusion of the aforementioned candidates, we have inserted some other important candidates like Vela X-1 \cite{gango, rawls}, 4U 1820-30 \cite{guver}, LMC X-4 \cite{rawls} and SMC X-1 \cite{aziz} to our analysis.

From Figs. \ref{fig1} and \ref{fig2}, together with Tables \ref{table1} and \ref{table2}, we have observed that for a mass range 1.04$\pm$0.09-2.67 $M_\odot$, the predicted radius lies on the range $9.79^{+0.26}_{-0.28}$-14.54 $km$ for the MIT Bag model and for the CFL phase of quark matter for the same mass range the radius falls in the range $9.14^{+0.24}_{-0.26}$-13.35 $km$. Upon comparison of these two EoSs and observational results, it is revealed that the CFL phase state is more realistic than the stiffer MIT Bag model. In the stiffer MIT Bag model, the existence of most of the candidates is ruled out. From Fig.s \ref{fig1} and \ref{fig2} in conjunction with Tables \ref{table1} and \ref{table2}, we have noticed that the radius of star candidates is increasing significantly while one considers the more positive value of the parameter $\eta$. This behaviour is noticed for both the considered cases of the EoSs.

For the considered pressure-density relations, the variations of maximum mass with the Rastall parameter $\eta$ are shown in Fig. \ref{fig3}. Looking at these figures, it is conjectured that maximum mass increases significantly once more negative $\eta$ parameters are approached for both EoSs. For the second EoS, however, a slight increase in maximum mass value is observed near the large value of $\eta$. The variation of radius $R$ corresponding to the maximum obtained mass ($M_\odot$) with $\eta$ are shown in Fig. \ref{fig4}. $\eta=0.2$ is corresponding to stellar structures with maximum radius and it is decreasing with decreasing $\eta$ values for both the considered cases of the present study. {It is interesting to note here that for the stiffer MIT Bag model EoS with more negative $\eta$ value ($\eta=-0.2$) the radius is slightly increasing. So for the radius variation, a non-linearity is observed.}

\subsection{Surface-redshift of stars}
Our primary concern in this section is to discuss the stability of the compact strange star model. In this regard, an important parameter is the surface redshift of stars. It concerns with the compactness of the stellar model. As mentioned in earlier sections, the considered star models are isotropic in nature, described with the perfect fluid matter. For such stars, the redshift can be defined as 
\begin{equation}
\label{a1}
Z=\frac{1}{\sqrt{(1-2u)}}-1,
\end{equation}
here the term $u(r)$ is the compactification factor defined as
\begin{equation}
\label{z1}
u(r)=\frac{m(r)}{r}.
\end{equation}
For a stable isotropic stellar model, this surface redshift parameter should be $Z\leq 2$ \cite{buchdahl, baraco}. Violation of this criterion will eventually lead to stellar structures with instabilities. 

\begin{figure*}
      	\centering{\includegraphics[width=7.5cm,height=6.5cm]{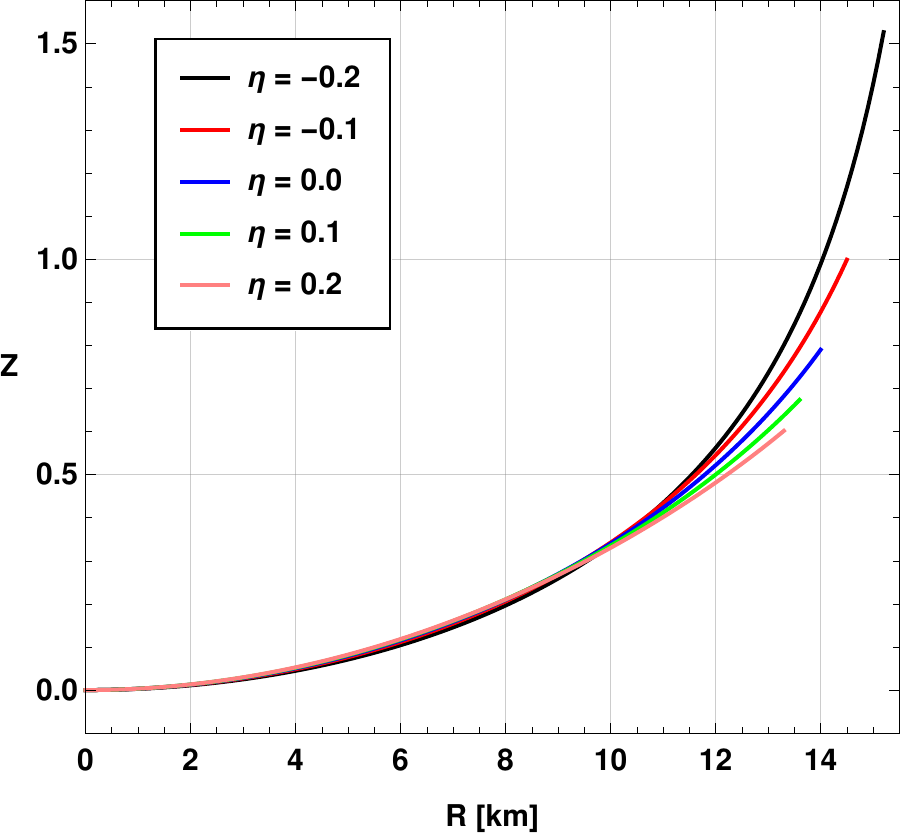}~~~~
      	\includegraphics[width=7.5cm,height=6.5cm]{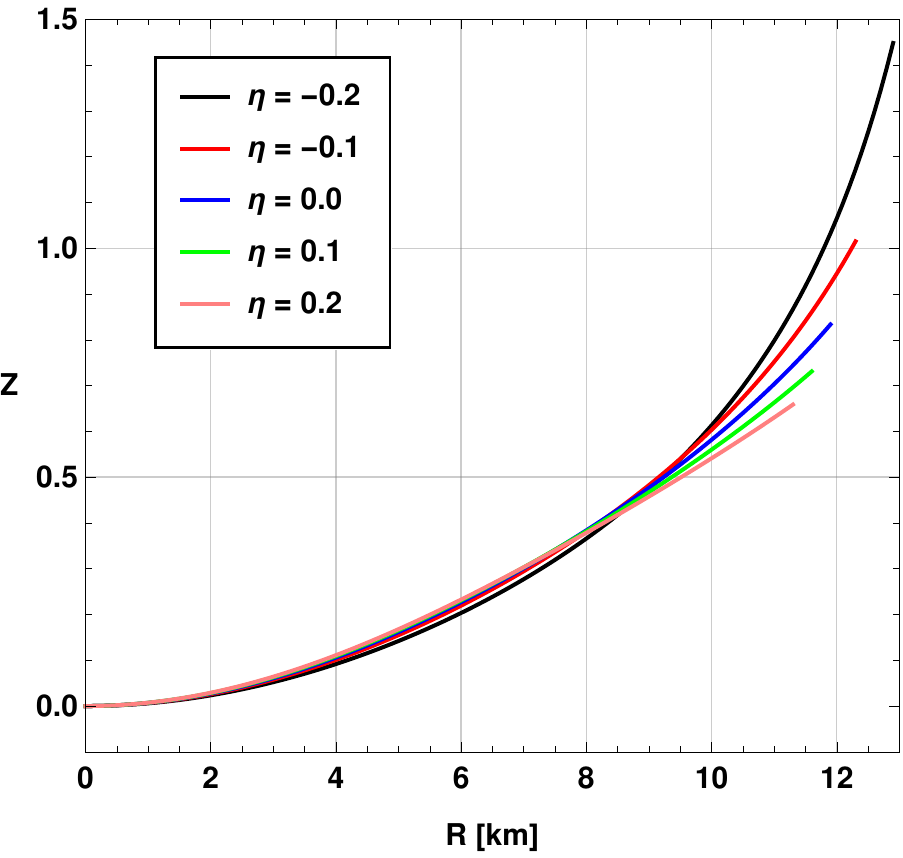}}
      	\caption{Variation of the surface redshift of compact stars with radial distance for MIT Bag model with $B = {(145\,\mbox{MeV})}^4$ (top panel) and CFL Phase state for $m_s=100 MeV$, $B = {(180\,\mbox{MeV})}^4$ and $\Delta = 470 MeV$ (bottom panel).}
      	\label{fig5}
            \end{figure*}

The variations of redshift with the radial distance for the strange star models with different $\eta$ values are shown in Fig. \ref{fig5}. The maximum surface redshift corresponds to $\eta=-0.2$ and the minimum corresponds to $\eta=0.2$, irrespective of the EoS chosen. With the increasing radial distance, the redshift is also increasing and it attains 
a maximum value at the surface of the star. In the first panel of this figure, the plots for the MIT Bag model are shown. From this plot, we have noticed that the maximum surface redshift is less than the limiting value of stable stellar structures. Similar conclusions can also be drawn by looking at the second panel of this figure where the surface redshifts are shown for the CFL phase state. In both our considered cases of EoSs and for all parametrization of the Rastall gravity model parameter $\eta$, the stable nature of stellar models is obtained from redshift analysis.

\subsection{Adiabatic index}
From the study of the adiabatic index of a stellar structure, one can comment on the stability of that star. Adiabatic stability criterion to check a region of stability of relativistic isotropic fluid spheres was derived earlier by \cite{1964a,1964b} and \cite{Ruffini}. 
The adiabatic index can be defined as \cite{chan}
\begin{equation}
\label{a2}
\Gamma=\frac{p+\rho}{p}\,\frac{dp}{d\rho}.
\end{equation}
It is shown that for the stability of a stellar structure $\Gamma > 4/3$ \cite{1964a,1964b,Ruffini}. To describe such stability one can define critical adiabatic index $\Gamma_{critical}$ \cite{mtw,1964a,1964b,mous} as \begin{equation}
\Gamma_{critical}=\frac{4}{3}+\alpha_0 \frac{M}{R}
\end{equation}
Here $\alpha_0$ is a small positive quantity, and $M/R$ is the compactness of the star. In the absence of the relativistic effect, the Newtonian limit can be regained from the above expression.

It also directly follows from the radial oscillations of a spherical star \cite{mtw}. The stability of spherical stars can be ensured when the fundamental mode of radial oscillations is a real quantity \cite{mtw}. For stability against radial perturbation $\Gamma>\Gamma_{critical}$ \cite{mous}. For stability of an isotropic sphere, adiabatic index $\Gamma > 4/3$ and violating this criterion will eventually lead to stellar instability. 

\begin{figure*}
      	\centering
       \includegraphics[width=7.5cm,height=6.5cm]{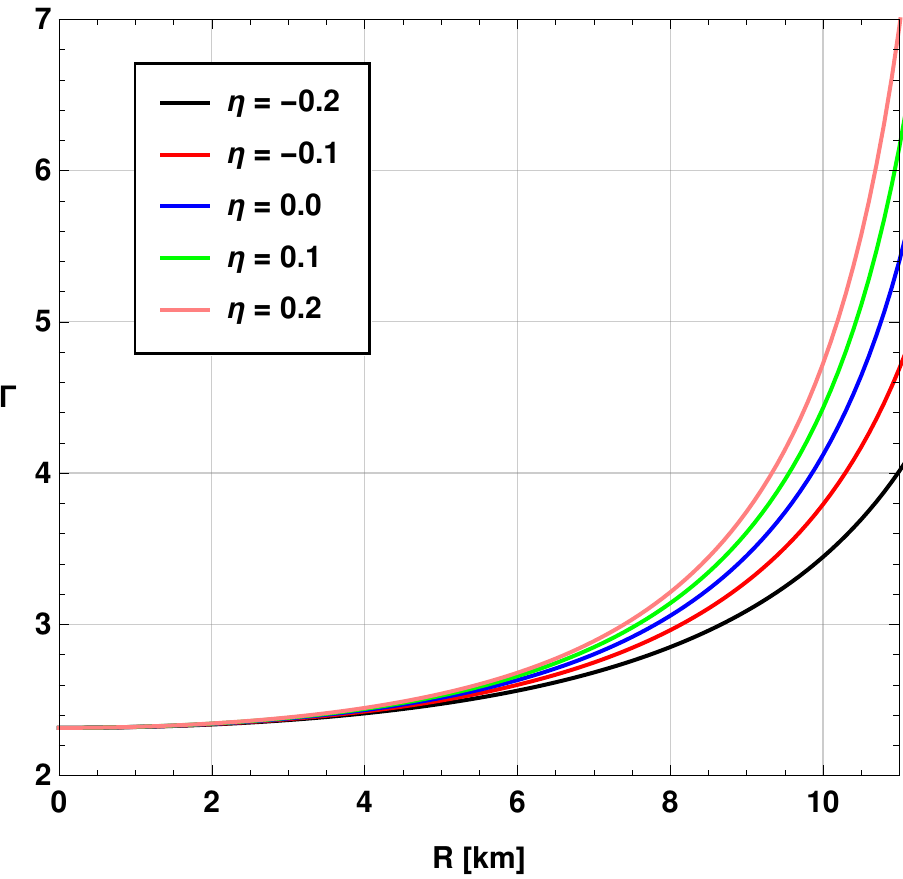}~~~~
      	\includegraphics[width=7.5cm,height=6.5cm]{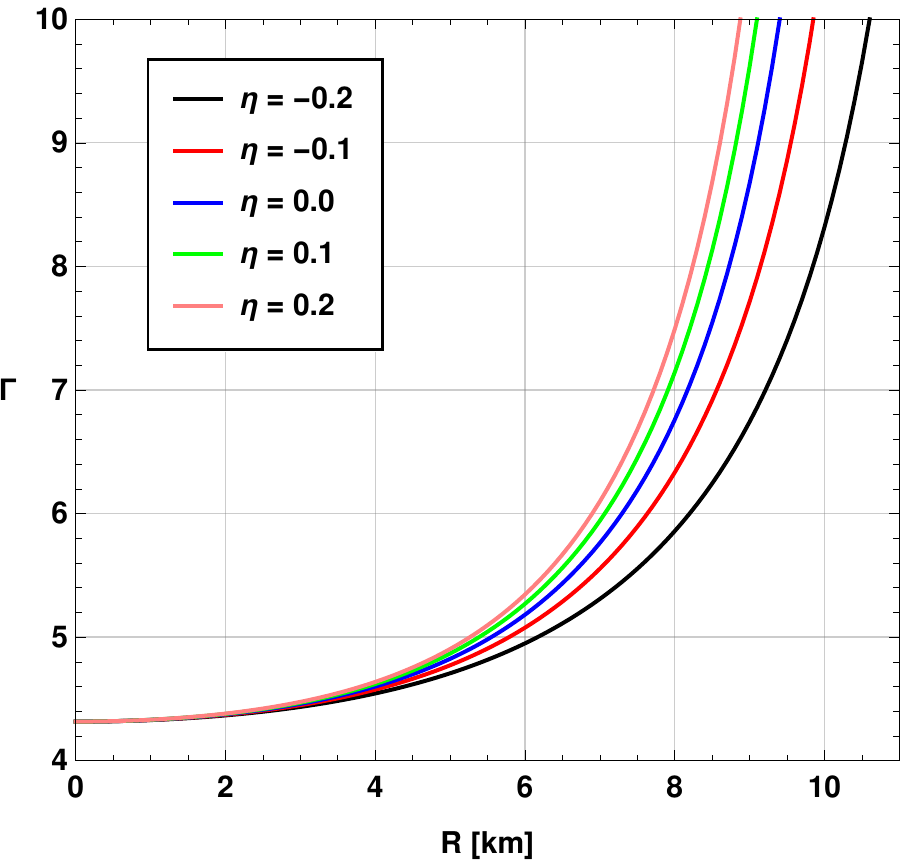}
      	\caption{Variation of the adiabatic index of compact stars along the radial distance for different $\eta$ values for strange stars with MIT Bag model with $B = {(145\,\mbox{MeV})}^4$ (top panel) and the CFL Phase state for $m_s=100 MeV$, $B = {(180\,\mbox{MeV})}^4$ and $\Delta = 470 MeV$ (bottom panel).}
      	\label{fig6}
            \end{figure*}

Looking at Fig. \ref{fig6}, it can be ensured that in our cases, the adiabatic index respects the Chandrasekhar stability criterion. In both EoSs, the adiabatic index $\Gamma$ increases along the radial distance. The minimum value of $\Gamma$ is obtained near the centre of each stellar model, and throughout the stellar model for each case, this value is greater than 2. Hence stability can be attained for these models.

            \begin{figure*}
      	\centering
            {\includegraphics[width=8cm,height=6.5cm]{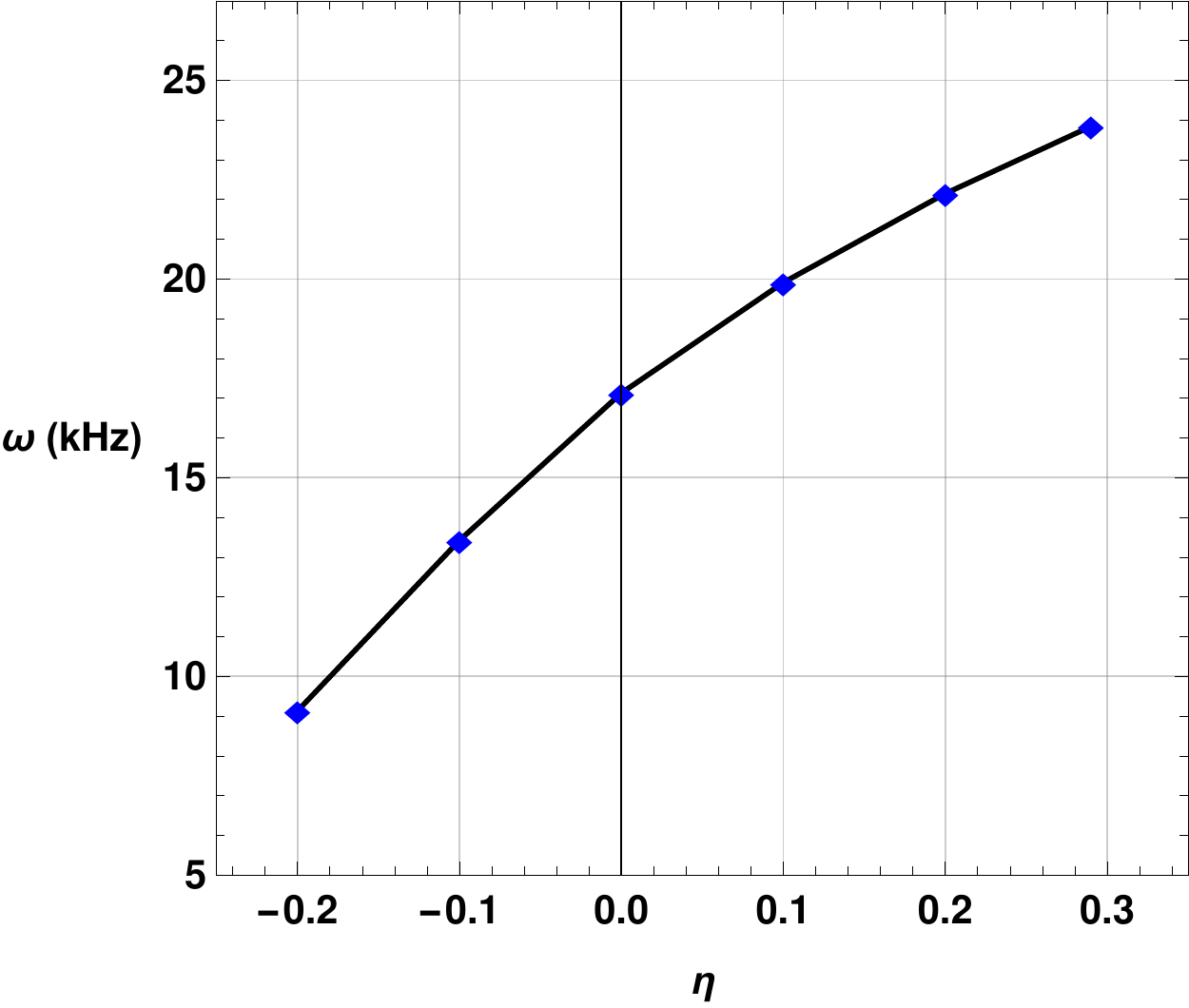}~~~~
      	\includegraphics[width=8cm,height=6.5cm]{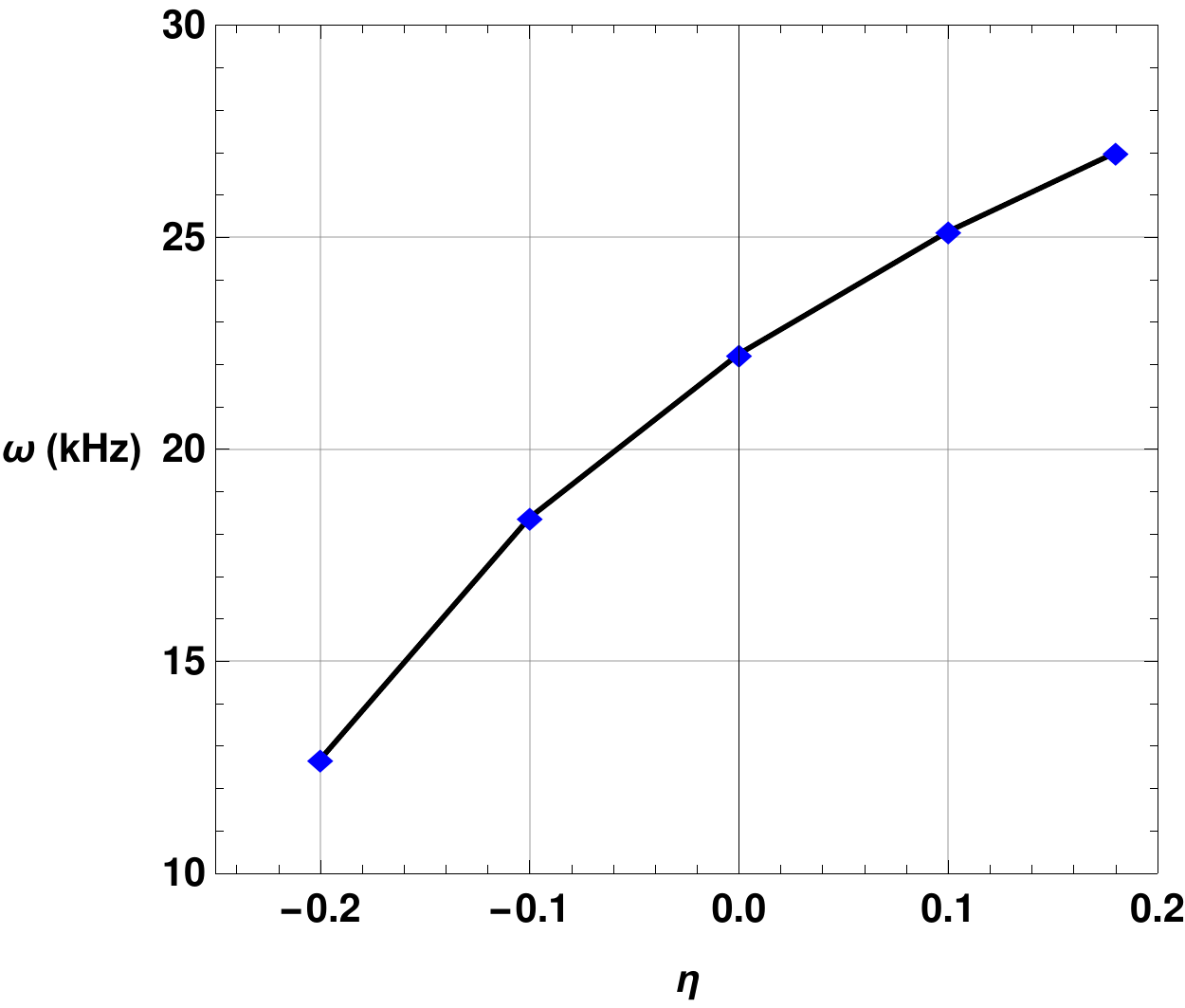}}
      	\caption{Variation of echo frequencies emitted from compact stars with Rastall parameter $\eta$ in MIT Bag model with $B = {(145\,\mbox{MeV})}^4$ and the CFL Phase state for $m_s=100 MeV$, $B = {(180\,\mbox{MeV})}^4$ and $\Delta = 470 MeV$.}
      	\label{fig7}
            \end{figure*}
\begin{table*}
{\centering
\begin{center}
\caption{\label{tab3} Parameters of strange stars for the MIT Bag model.}\vspace{2mm}
\scalebox{0.9}{\begin{tabular}{cccccccc}\hline
$\eta$ & Radius (R) & Mass (M) & Compactification  & GW echo  &  Echo & Surface & Adiabatic\\[-2pt] 
 & (in km) & (in $M_{\odot}$) & factor (M/R) & frequency (kHz) & time (ms) & redshift (Z) & index ($\gamma_{min}$)\\ \hline 
-0.2   & 13.56 & 3.55  & 0.3874 &  9.13  & 0.344 & 1.52 & 2.69\\
-0.1   & 13.58 & 3.38  & 0.3677 &  13.42 & 0.234 & 0.99 & 2.69\\
0      & 13.77 & 3.29  & 0.3540 &  17.13 & 0.183 & 0.79 & 2.69 \\ 
0.1    & 14.03 & 3.26  & 0.3438 &  19.91 & 0.158 & 0.67 & 2.69\\
0.2    & 14.34 & 3.26  & 0.3359 &  22.15 & 0.142 & 0.60 & 2.69\\ \hline
\end{tabular}}
\end{center}
}
\end{table*}

\begin{table*}
{\centering
\begin{center}
\caption{\label{tab4} Parameters of strange stars with CFL phase}\vspace{2mm}
\scalebox{0.9}{\begin{tabular}{cccccccc}\hline
$\eta$ & Radius (R) & Mass (M) & Compactification   & GW echo  & Echo & Surface & Adiabatic \\[-2pt] 
 & (in km) & (in $M_{\odot}$) & factor (M/R) & frequency (kHz) & time (ms)& redshift (Z) & index ($\gamma_{min}$)\\ \hline 
-0.2   & 12.00 & 3.03  & 0.3734 & 12.69 & 0.247 & 1.45 & 4.69 \\
-0.1   & 12.12 & 2.92  & 0.3562 & 18.39 & 0.171 & 1.01 & 4.69\\
0      & 12.36 & 2.88  & 0.3446 & 22.25 & 0.141 & 0.83 & 4.69\\ 
0.1    & 12.65 & 2.87  & 0.3355 & 25.14 & 0.124 & 0.73 & 4.69\\
0.2    & 12.96 & 2.89  & 0.3297 &  -     & -    & 0.66 & 4.69\\ \hline
\end{tabular}}
\end{center}
}
\end{table*}

\section{Gravitational wave echoes}\label{sec06}
As discussed above, after the GW observations, the knowledge of black holes and compact stars has taken a new turn. The discovery of binary black hole merger events prompted many researchers to investigate other exotic compact objects that potentially imitate black holes. In contrast to black holes, these compact objects can continue to exist with a high degree of compactness without having an event horizon. It was previously suggested that due to their high compactness, such compact objects might produce echoes of GWs hitting their gravitational potential barrier \cite{2018_Pani}. When GWs from a distant merger event strike their surface, the photon sphere reflects them, and after some delay, multiple reflections and refractions take place. Such objects must have a photon sphere at $R_p=3M$, where $M$ is the entire mass of the star, to produce echoes of GWs. Again, since the absence of an event horizon is one of the characteristics that distinguish compact stars from black holes, the minimum radius of such a star should be larger than Buchdahl's radius $R_b=9/4 M$. One may note that this Buchdahl's limits only apply to stars in GR considerations \cite{buchdahl}. In modified or alternative gravity theories, this value is modified to $R_b=9/(4-3c/2)M$ \cite{burikham}. Here $c=4\pi p_{\text{eff}}(R)R^2$. The term $p_{\text{eff}}(R)$ represents the effective pressure at the stellar surface. Again as the value of $c$ is negative and hence even for a small value of $c$ the Buchdahl's radius will decrease in the case of ATGs \cite{burikham,jb3}. Therefore, the compact stars that have radii that are within the range $R_b \leq R \leq R_p$ are the most likely to echo GWs striking their stellar surfaces.\\

As mentioned earlier, we have calculated the characteristic echo frequencies and echo times for the considered stellar models. To determine the echo frequencies, characteristic echo times are first obtained using the relation, 
\begin{equation}
	\label{z}
	\tau_{echo}\equiv\int_0^{3M}\!\!\!\! e^{\,(\lambda(r)-\chi(r))/2}\;dr.
	\end{equation} 
 
The metric functions $\lambda(r)$ and $\chi(r)$ can be evaluated using the metric \ref{eq1}. After calculating the echo time the echo frequencies can be calculated using the relation $\omega_{echo} \approx \pi/\tau_{echo}$ \cite{jb3,jb4}.\\

The obtained echo frequencies for the considered stellar model are shown in Fig. \ref{fig7}. In the first panel, the results are shown for the MIT Bag model and the other panel stars with the CFL phase state are shown. From these two figures, we can conjecture that, with an increase in $\eta$ values, a growth in frequencies is noticed. This variation is found to be the same for the two considered EoSs. Stars with CFL phase state emit echo frequencies with smaller values than the MIT Bag model EoS.  
\begin{table*}{
{\centering
\begin{center}
\caption{\label{tab5}Comparision of GR and Rastall theory (for $\eta=-0.2$)}\vspace{2mm}
\scalebox{0.9}{
\begin{tabular}{c|cccc|cccc}\hline
Gravity theory & & &Stiffer MIT Bag model & & &&CFL Phase&\\ \hline
 & Radius (R) & Mass (M) & M/R & GW echo   & Radius (R) & Mass (M) & M/R   & GW echo    \\[-2pt] 
 & (in km) & (in $M_{\odot}$) &  & frequency (kHz) & (in km) & (in $M_{\odot}$) &  & frequency (kHz) \\ \hline 
GR  & 13.77 & 3.29  & 0.3540 & 17.13 & 12.36 & 2.88 & 0.3446 & 22.25\\
Rastall & 13.56 & 3.55  & 0.3874 & 9.13  & 12.00 & 3.03 & 0.3734 & 12.69\\
 \hline
\end{tabular}}
\end{center}}}
\end{table*}

All the results discussed in the above sections are tabulated in Table \ref{tab3} and \ref{tab4}. In Table \ref{tab3}, strange stars with MIT Bag models are summarized, where we have added the numerical results for the five different values of the model parameter $\eta$. The mass of the most stable configuration of the mass-radius curve, corresponding radius, and compactness are shown together with other parameters like GW echo frequencies, echo time, surface redshift, and adiabatic index. 

For stars with the CFL phase state are listed in Table \ref{tab4}. One may note that for the Rastall parameter $\eta = 0.2$ for the case of CFL EoS, the compactness is around $0.3297$ which is too small to produce GW echoes. Hence a large positive value of the Rastall parameter is responsible for not producing GW echoes in such configurations.

\section{Concluding Remarks}
In this investigation, we have considered Rastall gravity, in which general energy-momentum conservation is violated, and studied strange stars explicitly. We have considered the modified TOV equations to study the mass-radius curves for two different EoSs {\it viz.,} MIT Bag model EoS and CFL EoS. We found that the energy-momentum conservation violation has a significant impact on the stellar structures. For negative values of the Rastall parameter $\eta$, the compactness increases, and the radius corresponding to the maximum mass in a mass-radius curve decreases (see Fig. \ref{fig4}). For $\eta>0$, the stellar configurations are less compact. This behaviour impacts the echo frequencies of GWs from such compact stars. For both EoSs, one can see that with an increase in the value of $\eta$, the GW echo frequency increases non-linearly. For the considered set of EoS parameters, we have found that the MIT Bag model can result in GW echoes with comparatively higher frequencies. For both EoSs the GW echo frequencies are found to be in the range of {$\approx 9-27$ kHz}. The surface redshifts for both the EoSs are affected by the energy-momentum conservation violation. With an increase in the Rastall parameter, the redshifts for both cases decrease non-linearly. {Furthermore, we have constrained the upper limit of Rastall parameter $\eta$ based on the compactification factor i.e., the emittance of echoes of GWs in this study. It is found that for the stiffer MIT Bag model, $\eta \leq 0.29$ and for the CFL phase EoS it is required that $\eta\leq 0.18$. For the two gravity theories, a comparison between the different results of parameter values is shown in Table \ref{tab5}. In this table, the values of mass, radius, compactness and echo frequency of strange stars corresponding to stiffer MIT Bag model and CFL Phase in GR and Rastall theory (with $\eta=-0.2$) are shown. For negative Rastall parameter, we have obtained stars with smaller radii, and more massive (and hence more compact) configurations for the two EoSs than that of the GR cases. The frequencies of GW echoes are also lower in Rastall theory with the chosen $\eta$ value. The difference between the radius, mass and echo frequencies as predicted by the Rastall theory (with $\eta=-0.2$) and GR for the stiffer MIT Bag model is about 0.26 km, 0.26$M_\odot$ and 8 kHz respectively. Whereas, the quantitative difference between that of compactness is about 0.03. For the case of CFL phase state, we have noticed a difference of about 0.36 km, 0.15$M_\odot$, 0.029 and 9.56 kHz respectively for radius, mass, compactness and echo frequency for the two theories. So,  from this study, it is also inferred that the assumption regarding the equivalence of Rastall's theory to Einstein's theory is refuted as we have noticed considerable deviations in the physical properties of the considered compact stars for both the EoSs. Furthermore, for more negative Rastall parameter $\eta$ we have noticed echo frequency comes closer to the detectable range of current GW detectors \cite{martynov}.
It is important to note here that the GW merging events GW170817 and GW190425 have indicated no tidal deformability in GW signals and it is inferred that the upper bound is compatible with soft EoSs. In the present study, the stiffer MIT Bag model is only used to obtain stars with large compactness which can emit echoes of GWs as from soft EoSs one can't expect a star which can echo the GW. Again, we have inferred that rather than using these ad-hoc stiffer EoS, the echo possibilities can be expected from more realistic CFL phase EoS. For negative Rastall parameter values the obtained echo frequencies are much smaller than that obtained from GR case.}
In the near future, with the help of a next-generation GW detector like LISA, it might be possible to have a stringent constraint on the energy-momentum conservation violation or the Rastall parameter. 

{It is worth mentioning here some recent studies on quark stars in Rastall theory of gravity. Recently, in Ref. \cite{2023_Banerjee} the authors have chosen quantum chromodynamics motivated EoS to study the properties of quark stars. It is also reported therein that the value of the maximum mass is more than $2M_{\odot}$. In Ref. \cite{Tangphati}, the authors have reported on anisotropic quark star models in Rastall gravity considering MIT Bag model EoS. Whereas in this present work, we have studied only the isotropic strange stars using two EoSs: the stiffer MIT Bag model and CFL phase EoS. Furthermore, in this study, we have considered the possibilities of the emission of echoes of GWs from strange stars considering the Rastall theory of gravity. To the best of our knowledge, echo possibilities from strange stars are not considered earlier for the Rastall theory of gravity. }

Such a study can be further extended to Rastall--$f(R)$ gravity theory, Rastall--Gauss-Bonnet theory, etc. with other promising EoSs. Apart from strange stars, similar studies can be extended to gravastars, white dwarfs, etc. to understand more about the impacts of energy-momentum conservation violation on stellar structures. Reconstruction of a more realistic EoS also can be another promising aspect of study in theories with energy-momentum conservation violation.

\section*{Acknowledgments}
JB and DJG are thankful to Prof. U. D. Goswami for some useful discussions. DJG acknowledges the contribution of the COST Action CA21136  -- ``Addressing observational tensions in cosmology with systematics and fundamental physics (CosmoVerse)". G. Mustafa acknowledges Grant No. ZC304022919 to support his Postdoctoral Fellowship at Zhejiang Normal University. The author SKM is thankful for continuous support and encouragement from the administration of University of Nizwa.

\section*{Data Availability Statement}
There are no new data associated with this article.


\begin{thebibliography}{99}
\bibitem{prl116}
B. P. Abbott et al., [The LIGO Scientific Collaboration and The Virgo Collaboration], Phys. Rev. Lett. 116, 061102 (2016).

\bibitem{prl118} B. P. Abbott et al. [LIGO Scientific and VIRGO], Phys.
Rev. Lett. 118, 22, 221101 (2017) [erratum: Phys.
Rev. Lett. 121, 12, 129901 (2018)]

\bibitem{prl119}
B. P. Abbott et al. [LIGO Scientific and Virgo], Phys.
Rev. Lett. 119, no.16, 161101 (2017)

\bibitem{prx9}
B. P. Abbott et al. [LIGO Scientific and Virgo], Phys.
Rev. X 9, 3, 031040 (2019)

\bibitem{gw190814}
R. Abbott, T. D. Abbott, S. Abraham, S., et al. [The LIGO Scientific Collaboration and The Virgo Collaboration], ApJL, 896, L44 (2020)

\bibitem{Zhang}
C. Zhang, Phys Rev D 104 083032 (2021)

\bibitem{Zhang2}
C. Zhang, Y. Gao, C. Xia, \& R. Xu, arXiv:2305.13323 (2023)

\bibitem{jb4}
J. Bora, J. U.D. Goswami, Physics of the Dark Universe  38, 101132 (2022)

\bibitem{2018_Pani}
P. Pani, \& V. Ferrari, Class. Quantum Grav. 35, 15LT01 (2018)

\bibitem{2018_manarelli}
M. Mannarelli \& F. Tonelli, Phys Rev D 97, 123010 (2018)

\bibitem{2019_Urbano}
A. Urbano, \& H. Veerm\"ae, J Cosmol Astropart Phys, 2019, 011 (2019)

\bibitem{jb1}
J. Bora, U.D.Goswami, Monthly Notices of the Royal Astronomical Society 502, 1557, (2021)

\bibitem{jb2}
J. Bora, \& U.D. Goswami, Astropart. Phys. 143, 102744 (2022a).

\bibitem{jb3}
J. Bora, D.J. Gogoi \&  U.D.Goswami, J. Cosmol. Astropart. Phys. 09, 057 (2022)

\bibitem{0451}
M.C. Miller, F.K. Lamb, A.J.Dittmann et al. ApJ, 887, L24 (2019)

\bibitem{0451_2}
T.E. Riley, A.L. Watts, S. Bogdanov et al., ApJ, 887, L21 (2019)

\bibitem{6620_1}
M.C. Miller, F.K. Lamb, A.J. Dittmann et al. ApJL, 918, L28 (2021)

\bibitem{6620_2}
T.E. Riley, A.L. Watts, P.S. Ray et al. ApJL, 918, L27 (2021)

\bibitem{romani}
R.W. Romani, D. Kandel, A.V. Filippenko, T.G. Brink \& W. Zheng, ApJL, 934, L17 (2022)

 
\bibitem{Tangphati1}
T. Tangphati, A.Pradhan, A. Errehymy, \& A. Banerjee, Physics Letters B, 819, 136423 (2021)

\bibitem{gogoi1}
D.J. Gogoi \& U.D. Goswami, Eur. Phys. J. C 80, 1101 (2020)

\bibitem{gogoi_cosmo}
D.J. Gogoi \& U.D. Goswami, Int. J. Mod. Phys. D 31, 2250048 (2022)

\bibitem{gogoi4}
D.J. Gogoi, \& U.D. Goswami, Indian J. Phys., 96, 637 (2022)

\bibitem{Koussour23A}
M. Koussour, S. Arora, D.J. Gogoi, M. Bennai, \& P.K. Sahoo,  Nuclear Physics B, 990, 116158 (2023)

\bibitem{Koussour23B}
M. Koussour,  \& D.J. Gogoi, preprint [arXiv:2303.04640] (2023)

\bibitem{sk19}
S.K. Maurya, A. Errehymy, D. Deb,F.  Tello-Ortiz, \& M. Daoud, Phys Rev D, 100, 044014 (2019)

\bibitem{gogoi5}
D.J. Gogoi, \& U.D. Goswami, J Cosmol Astropart Phys 02, 027 (2023c)

\bibitem{23_sk}
S.K. Maurya, K.N. Singh, M. Govender, \& S.  Ray, Monthly Notices of the Royal Astronomical Society, 519, 4303 (2023)

\bibitem{Lambiase23}
G. Lambiase, R.C. Pantig, D.J. Gogoi, \& A. \"Ovg\"un,  preprint [arXiv:2304.00183] (2023)

\bibitem{GogoifQ}
D.J. Gogoi, A. \"Ovg\"un, \& M. Koussour, preprint [arXiv:2303.07424] (2023)

\bibitem{Sekhmani23}
Y. Sekhmani, \& D.J. Gogoi, Int. J. Geom. Methods Mod. Phys., S0219887823501608 (2023)

\bibitem{gogoi3}
D.J. Gogoi \& U.D.Goswami, Physics of the Dark Universe 33, 100860 (2021)

\bibitem{gogoi2}
D.J. Gogoi, \& U.D. Goswami, J Cosmol Astropart Phys, 06, 029 (2022)

\bibitem{2021_Silva}
F.M. Da Silva, L.C.N. Santos, \& C.C. Barros, Class Quantum Grav, 38, 165011 (2021)

\bibitem{2021_Salako}
I.G.Salako, D.R. Boko, G.F. Pomalegni \& M.Z. Arouko, Theor Math Phys, 208, 1299 (2021)

\bibitem{2021_Ghosh}
S. Ghosh, S. Dey, A. Das, A. Chanda,  \& B.C. Paul, J Cosmol Astropart Phys, 2021, 004 (2021)

\bibitem{2019_Maulana}
H. Maulana, \& A. Sulaksono, Phys Rev D 100, 124014 (2019)
\bibitem{Astash01}  A. V. Astashenok , S. Capozziello,  S. D. Odintsov, \& V. K. Oikonomou, Phys. Lett. B, 811, 135910 ( 2020) 

\bibitem{Astash1}  A. V. Astashenok , S. Capozziello,  S. D. Odintsov, \& V. K. Oikonomou,  Phys. Lett. B, 816, 136222  (2021)

\bibitem{rastall}
P. Rastall, Phys Rev D, 6, 3357 (1972)

\bibitem{23_Gogoi2}
D.J. Gogoi, R. Karmakar, \& U.D. Goswami, Int J Geom Methods Mod Phys, 20, 2350007, (2023)

\bibitem{1971_Bodmer}
A.R. Bodmer, Phys Rev D, 4, 1601 (1971)

\bibitem{1984_Witten}
E. Witten, Phys Rev D  30, 272 (1984)

\bibitem{1986_Alcock} 
C. Alcock,  E. Farhi,  \& A. Olinto,  Apj  310, 261 (1986)

\bibitem{Weber1999} 
F. Weber, R. Negreiros, P. Rosenfield,  \& M. Stejner, Progress in Particle \& Nuclear Physics, 59, 94 (2007)
{
\bibitem{Moradpour17}
H. Moradpour, N. Sadeghnezhad, \& S. H. Hendi, Can. J. Phys. 95, 1257 (2017)
\bibitem{morad}
H. Moradpour, Y. Heydarzade, F. Darabi, \& I. G. Salako, Eur. Phys. J. C 77, 259 (2017)
\bibitem{Koivisto}
T. Koivisto, Class. Quant. Grav. 23, 4289 (2006)
\bibitem{Bertolami}
O. Bertolami, C. G. Boehmer, T. Harko \& F. S. N. Lobo, Phys. Rev. D 75, 104016 (2007)
\bibitem{Harko}
T. Harko, \& F. S. N. Lobo, Galaxies, 2, 410 (2014)
\bibitem{rawaf1}
A. S. Al-Rawaf, \& O. M. Taha, Phys. Lett. B 366, 69 (1996)
\bibitem{rawaf2}
A. S. Al-Rawaf, \& O. M. Taha, Gen. Rel. Grav. 28, 935 (1996)
\bibitem{Li}
R. Li, J. Wang, Z. Xu, \& X. Guo, Mon. Not. Roy. Astron. Soc. 486, 2407 (2019)
\bibitem{rawaf3}
A. S. Al-Rawaf, Int. J. Mod. Phys. D 14, 1941 (2005)
\bibitem{Majernik}
V. Majernik, Gen. Rel. Grav. 35, 1007 (2003)
\bibitem{Arbab}
A. I. Arbab, JCAP. 05, 008 (2003)
\bibitem{Rahman1}
A.-M. M. Abdel-Rahman, Astrophys. Space Sci. 278, 383 (2001)
\bibitem{Rahman2}
A.-M. M. Abdel-Rahman, \& M. H. A. Hashim, Astrophys. Space Sci. 298, 519 (2005)
\bibitem{ziaie}
A. H. Ziaie, H. Moradpour \& S. Ghaffari, Physics Letters B 793, 276 (2019)
\bibitem{visser}
M. Visser, Phys.Lett. B, 782, 83 (2018)
\bibitem{Darabi}
F. Darabi, H. Moradpour, I. Licata, Y. Heydarzade, \& C. Corda, Eur. Phys. J. C 78, 25 (2018)}

\bibitem{2015_Oliveira}
A.M. Oliveira, H.E.S. Velten, L. Casarini,  \&J.C. Fabris, Phys Rev D 92, 044020  (2015)

\bibitem{2023_Banerjee}
A. Banerjee, T. Tangphati \& A. Pradhan, Int. J. Mod. Phys. D, 32,2350026  (2023)

\bibitem{2022_mustafa}
G. Mustafa, A. Errehymy, A. Ditta, \& M. Daoud, Chinese Journal of Physics 77, 2781 (2022)

\bibitem{2021_mustafa}
G. Mustafa, X. Tie-Cheng, \& M. F. Shamir, Phys Scr, 96, 105008 (2021)

\bibitem{zubair}
M. Zubair, S. Waheed, M.F. Jamal, \& G. Mustafa, Results in Physics, 29, 104787 (2021)

\bibitem{mustafa}
G. Mustafa, I. Hussain, M.F. Shamir, \& X. Tie-Cheng, Phys Scr 96, 045009 (2021b)
{{
\bibitem{abbas}
G. Abbas \& M. R. Shahzad, Eur. Phys. J. A 54, 211 (2018)
\bibitem{abbas2}
G. Abbas \& M. R. Shahzad, Astrophys Space Sci 364, 50 (2019)
}}

\bibitem{20_sk}
S.K. Maurya, \& F. Tello-Ortiz, Physics of the Dark Universe 29, 100577 (2020)

\bibitem{21_gogoi}
D.J. Gogoi \& U. Dev Goswami,Physics of the Dark Universe, 33, 100860 (2021)

\bibitem{23_gogoi}
D.J. Gogoi, Y. Sekhmani, D. Kalita, N.J. Gogoi, \& J. Bora, Fortschritte der Physik, 2300010, (2023)

\bibitem{2019_Shahzad}
M.R. Shahzad,  \& G. Abbas, Int J Geom Methods Mod Phys, 16, 1950132 (2019)

\bibitem{22_sk}
S.K. Maurya, Ksh. Newton Singh, M. Govender, \& S. Hansraj, ApJ, 925, 208 (2022)

\bibitem{2023_kumar}
A. Kumar, V.B. Thapa, \& M.Sinha,  Phys Rev D, 107 (2023)

\bibitem{tolman}
R.C. Tolman, Phys Rev, 55, 364 (1939) 

\bibitem{oppenheimer}
J.R. Oppenheimer, \& G.M. Volkoff, Phys Rev, 55, 374 (1939)

\bibitem{weinberg1972gravitation}
S. Weinberg, Gravitation and cosmology: principles and applications of the general theory of relativity (New York: Wiley) (1972)

\bibitem{glendenning2000compact}
N.K. Glendenning,  Compact Stars: Nuclear Physics, Particle Physics, and General Relativity, 2nd ed (Springer, New York) (2000)
{
\bibitem{Moradpour16}
H. Moradpour, \& I. G. Salako, Advances in High Energy Physics 2016, 3492796 (2016)
\bibitem{hanafy}
W. El Hanafy, ApJ 940, 51 (2022)
\bibitem{hanafy2}
W. El Hanafy \& A. Awad, ApJ 951, 144 (2023)
\bibitem{Tangphati}
T. Tangphati, A. Banerjee, S. Hansraj, \& A. Pradhan, Annals of Physics 452, 169285 (2023)
 \bibitem{Banerjee23}
A. Banerjee, T. Tangphati, S. Hansraj, \& A. Pradhan, Annals of Physics 451, 169267 (2023)
\bibitem{2017_Moradpour}
H. Moradpour, A. Bonilla, E. M. C. Abreu, \& J. A. Neto, Phys. Rev. D 96, 123504 (2017)}



\bibitem{alford}
M. Alford,M.  Braby, M. Paris \& S. Reddy, ApJ, 629,969 (2005)

\bibitem{lugones}
G. Lugones, \& J.E. Horvath, Phys. Rev. D  66, 074017 (2002)

\bibitem{flores}
C.V. Flores \& G. Lugones, Phys. Rev. C  95, 025808 (2017)

\bibitem{farhi}
E. Farhi \& R.L. Jaffe, Phys. Rev. D  30, 2379 (1984)

\bibitem{beringer}
J. Beringer et al. Phys Rev D, 86, 010001  (2012)

\bibitem{gango}
T. Gangopadhyay, S. Ray, X.-D. Li, J. Dey, \& M. Dey, MNRAS 431, 3216 (2013)

\bibitem{rawls}
M.L. Rawls, J.A. Orosz, J.E. McClintock et al., ApJ, 730, 25 (2011) 

\bibitem{guver}
T. G\"uver, P. Wroblewski ,L.  Camarota, \&  F. \"Ozel, ApJ, 719, 1807 (2010)

\bibitem{aziz}
A. Aziz, S. Ray, F. Rahaman, M. Khlopov, \& B.K. Guha, Int J Mod Phys D, 28, 1941006 (2019)

\bibitem{antoni}
J. Antoniadis,P.C.C. Freire, N. Wex et al. Science, 340, 1233232 (2013)

\bibitem{buchdahl}
H.A. Buchdahl, Phys Rev, 116, 1027 (1959)

\bibitem{baraco}
D. Barraco \& V.H. Hamity, Phys Rev D, 65, 124028 (2002)

\bibitem{1964a}
S. Chandrasekhar, ApJ, 140, 417 (1964a)

\bibitem{1964b}
S. Chandrasekhar, Phys Rev Lett, 12, 114 (1964b) 

\bibitem{Ruffini}
M. Merafina, \& R. Ruffini,  A\&A, 221, 4 (1989)

\bibitem{chan}
R. Chan, L. Herrera \& N.O. Santos, MNRAS, 265, 533 (1993)

\bibitem{mtw}
C.W. Misner, K.S. Thorne, \& J.A. Wheeler, Gravitation (San Francisco: W. H. Freeman) (1973)

\bibitem{mous}
Ch. C. Moustakidis,Gen Relativ Gravit, 49, 68 (2017)

\bibitem{burikham}
P. Burikham, T. Harko, \& M.J.Lake, Phys Rev D, 94, 06407 (2016)


\bibitem{martynov}
D. Martynov et al., Phy. Rev. D, 99, 102004 (2019)

\end{thebibliography}
\end{document}